\documentclass[12pt]{article}%
\usepackage[authoryear]{natbib}
\usepackage{amsfonts}
\usepackage{amsmath,amssymb,amsthm,enumerate,epsfig,graphicx}
\usepackage{ifthen,latexsym,syntonly}
\usepackage{natbib}
\usepackage{amsmath}
\usepackage{color}
\usepackage{setspace}
\usepackage{amssymb}
\usepackage{graphics}
\usepackage{graphicx}
\usepackage{subfigure}
\usepackage{amsthm}
\usepackage{mathrsfs}
\usepackage{lscape}
\usepackage{multirow}
\usepackage{floatrow}
\allowdisplaybreaks
\newtheorem{theorem}{Theorem}

\newtheorem{lemma}{Lemma}
\newtheorem{proposition}{Proposition}

\renewcommand{\cite}{\citet*}

\newcommand{\bb}{\mbox{\bf b}}
\newcommand{\bff}{\mbox{\bf f}}
\newcommand{\bx}{\mbox{\bf x}}

\newcommand{\bA}{\mbox{\bf A}}

\newcommand{\bB}{\mbox{\bf B}}
\newcommand{\bC}{\mbox{\bf C}}
\newcommand{\bD}{\mbox{\bf D}}

\newcommand{\bF}{\mbox{\bf F}}

\newcommand{\bI}{\mbox{\bf I}}

\newcommand{\bT}{\mbox{\bf T}}

\newcommand{\bU}{\mbox{\bf U}}
\newcommand{\bV}{\mbox{\bf V}}
\newcommand{\bX}{\mbox{\bf X}}
\newcommand{\bW}{\mbox{\bf W}}
\newcommand{\bY}{\mbox{\bf Y}}
\newcommand{\bZ}{\mbox{\bf Z}}
\newcommand{\bone}{\mbox{\bf 1}}

\newcommand{\bzero}{\mbox{\bf 0}}

\newcommand{\bmu}{\mbox{\boldmath $\mu$}}
\newcommand{\bgamma}{\mbox{\boldmath $\gamma$}}
\newcommand{\bepsilon}{\mbox{\boldmath$\epsilon$}}

\newcommand{\bnu}{\mbox{\boldmath$\nu$}}

\newcommand{\bSigma}{\mbox{\boldmath $\Sigma$}}

\newcommand{\brho}{\mbox{\boldmath $\rho$}}

\newcommand{\diag}{\mathrm{diag}}
\newcommand{\argmin}{\mathrm{argmin}}

\newcommand{\FDP}{\mbox{FDP}}
\newcommand{\argmax}{\mbox{argmax}}

\begin{document}

\title{\bf  Large-Scale Multiple Testing for Matrix-Valued Data under Double Dependency}

\author{ Xu Han, Sanat Sarkar, Shiyu Zhang}
\date{}

\maketitle

\pagestyle{myheadings} \normalsize

\begin{abstract}
\begin{singlespace}
High-dimensional inference based on matrix-valued data has drawn increasing attention in modern statistical research, yet not much progress has been made in large-scale multiple testing specifically designed for analysing such data sets. Motivated by this, we consider in this article an electroencephalography (EEG) experiment that produces matrix-valued data and presents a scope of developing novel matrix-valued data based multiple testing methods controlling false discoveries for hypotheses that are of importance in such an experiment. The row-column cross-dependency of observations appearing in a matrix form, referred to as double-dependency, is one of the main challenges in the development of such methods. We address it by assuming matrix normal distribution for the observations at each of the independent matrix data-points. This allows us to fully capture the underlying double-dependency informed through the row- and column-covariance matrices and develop methods that are potentially more powerful than the corresponding one (e.g., Fan and Han (2017)) obtained by vectorizing each data point and thus ignoring the double-dependency. We propose two methods to approximate the false discovery proportion with statistical accuracy. While one of these methods is a general approach under double-dependency, the other one provides more computational efficiency for higher dimensionality. Extensive numerical studies illustrate the superior performance of the proposed methods over the principal factor approximation method of Fan and Han (2017). The proposed methods have been further applied to the aforementioned EEG data.
\end{singlespace}
\end{abstract}
\noindent {{\bf Keywords:} matrix valued data, large scale multiple comparison, false discovery proportion, double dependences, electroencephalogram. }

\newpage

\section{Introduction}

Large-scale multiple testing is an integral part of statistical investigations in the modern era of Big Data-driven scientific research with statisticians/data scientists frequently encountering simultaneous testing of tens of thousands or even hundreds of thousands of hypotheses in such research. Despite substantial growth of research in multiple testing over the past few decades, development of multiple testing methods specifically designed for matrix-valued data has not yet received much attention, even though such data have been increasingly seen to occur in various applications, for instance, in brain imaging, electroencephalography (EEG), environmental, finance, economics and many others.

The row-column cross-dependency of observations appearing in a matrix-structured form at each data-point, which we refer to as double-dependency in this article, is a newer challenge in developing a multiple testing method specifically designed for matrix-valued data. One can, of course, mitigate this challenge by vectorizing each data-point and consider using an appropriately chosen method, depending on the problem under consideration, from the abundant literature on vector-valued data-based multiple testing methods. However, such a method does not utilize the original matrix structure of the data, and so would be less desirable and potentially less powerful than the one that can be developed by capturing the underlying double-dependency. So, there seems to be an urgent need for developing such matrix-valued data-based multiple testing methods.

Driven by the aforementioned need, we revisit the matrix-valued data set from an EEG experiment, used by Li, Kim and Altman (2010) and many other researchers while developing newer statistical theories and methodologies for such data sets (see also Nandi and Sarkar (2021)). This data set presents an opportunity for us to develop our desired novel multiple testing methods, at least in the context of such an important scientific investigation. The EEG experiment involved a control group and a treatment group comprising of alcoholic subjects. Ten trials were performed on each subject and a picture was presented to the subject during each trial, while EEG activity in the form of voltage fluctuations (in microvolts) were recorded at 256 time points from 61 electrodes placed on the subject's scalp. Figure 1 in Li et al. (2010) shows an example of the EEG pattern, averaged over measurements obtained from ten trials, for two subjects, one each from the control and alcoholic groups. This figure clearly indicates a difference in the voltage fluctuation patterns for the two subjects over time and electrodes, sparking our interest in developing novel multiple testing methods for comparing two groups that can potentially be applied to gain deeper understanding of brain dysfunction and regions impacted by alcoholism. It is important to note that we aim at developing such methods in the framework of thresholding estimated false discovery proportion (FDP) and, as indicated above, will be aimed at doing so by fully capturing the underlying double-dependency without characterizing it  through any condition. More specifically, we will extend the work of Fan and Han (2017) from vector-valued to matrix-valued data.


Substantial challenges do arise in developing multiple testing methods for matrix-valued data. First of all, the number of hypotheses is often excessively large, relative to the sample size, even when each dimension of the matrix is not very high, since the product of the two dimensions can be ``quadratically high". In the EEG data, for example, there are $64\times256=16384$ hypotheses to be tested when comparing the two groups. One naive way to handle this challenge would be to vectorize the matrix data by stacking the columns and apply the Principal Factor Analysis (PFA) method in Fan and Han (2017) to the vectorized data assuming vector-variate multivariate normal. The PFA relies on an estimate of the unknown covariance matrix, particularly through the eigenvalues and eigenvectors of this matrix. When the dimensionality is quadratically high and the sample size is relatively small, the performance of PFA in approximating the FDP will deteriorate. We will illustrate this issue in the numerical studies. The second challenge, as noted above, is the double-dependency and its effective full utilization into our methods.
We will handle this challenge by assuming that the underlying random matrix of voltage observations, say $\bX$, follows a matrix normal distribution; i.e., $\bX \sim \mathcal{MN}(\bmu_{p\times q},\bU_{p\times p}, \bV_{q\times q})$, where $\bmu_{p\times q}$ is the location matrix parameter, $\bU$ and $\bV$ are, respectively, the commong covariance matrices of the column and row vectors. The matrix normal is theoretically amenable to newer methodological developments specifically for matrix-valued data and so has been a commonly used distribution for analysing such data sets; see, i.e., Li, Kim and Altman (2010) and Xia and Li (2017), respectively, for the development of multivariate regression and dimension folding and for brain connection testing. The matrix normal has also been used for analysing microarray data (Allen and Tibshirani (2012)) and mRNA expression data (Hornstein, Fan, Shedden and Zhou (2019)). More importantly for our research, the underlying double-dependency can be effectively parameterized through this distribution.

In the current paper, we propose two methods for approximating the FDP based on matrix normal data. The first method, called the noodle method, utilizes the property of matrix normal distribution through Kronecker product. More specifically, the vectorized matrix normal, having a multivariate normal distribution with the covariance matrix as the Kronecker product of the row and column covariance matrices, provides a structural information about the underlying double-dependency, and thus provides a dimension reduction advantage. This not only allows a full-capture of the double-dependency but also facilitates estimation of FDP in a large-scale multiple testing setting. Instead of estimating a $(pq)\times(pq)$ dimensional covariance matrix for the vectorized data, we are actually estimating two smaller matrices: $p\times p$ dimensional column covariance matrix $\bU$ and $q\times q$ dimensional row covariance matrix $\bV$.

Although the noodle method shows superior performance for matrix data in comparison with the PFA procedure, it suffers from some computational complexity issues. More specifically, in the first method, we need to calculate any pair of the eigenvalues and eigenvectors from the two estimated covariance matrices, $\widehat{\bU}$ and $\widehat{\bV}$. When $p$ and $q$ are large, the noodle method is computational intensive. To circumvent this issue, we propose the second method, the sandwich method, which involves the first few principal components from $\widehat{\bU}$ and $\widehat{\bV}$, respectively, mostly capturing the underlying dependence structure. The sandwich method is developed to handle large number of tests, much larger than when the noodle method can be used. Our simulation studies will show that the sandwich method can be applied to a more ambitious setting where $p\times q=500\times 500=250000$, where the noodle method fails.

The rest of the paper is organized as follows. In Section 2, we describe the two proposed methods with theoretical justifications. Section 3 provides results of simulation studies we conducted to compare these methods with the PFA method in Fan and Han (2017) under various scenarios.  Section 4 presents the results obtained from the analysis of the EEG data using these methods. All technical proofs are relegated to Appendix.

\section{Main Results} Our proposed methods will be presented in this section. First, let us introduce below some of the notations to be used throughout this paper.
\begin{itemize}
\item [$\bullet$] $a_n\asymp b_n = 0<a_n/b_n+b_n/a_n=O(1)$. \

\item [$\bullet$] For a vector $\bx=(x_1,\cdots,x_p)'$, $\|\bx\|=\sqrt{\sum_{i=1}^px_i^2}$, $\|\bx\|_1=\sum_{i=1}^p|x_i|$. \

\item [$\bullet$] For a matrix $\bA=(a_{ij})\in\mathbb{R}^{m\times n}$, Frobenius norm: $\|\bA\|_F=\sqrt{\rm{trace}(\bA'\bA)}$; Operator norm: $\|\bA\|=\lambda_{\max}^{1/2}(\bA^T\bA)$; $l_1$ norm: $\|\bA\|_1=\max_{1\leq j\leq n}\sum_{i=1}^m|a_{i,j}|$; and $l_{\infty}$ norm: $\|\bA\|_{\infty}=\max_{1\leq i\leq m}\sum_{j=1}^n|a_{ij}|$. \

\item [$\bullet$] For two matrices $\bA_{n\times m} = (a_{ij})$ and $\bB_{p\times q}= (b_{ij})$,
\begin {eqnarray} \mbox{Kronecker product of} \; \bA \; \mbox{and} \; \bB &:& \bA \otimes \bB = \left(\begin{array}{ccc}a_{11}\bB & \cdots & a_{1m}\bB \\ \vdots &  & \vdots \\ a_{n1}\bB & \cdots & a_{nm}\bB\end{array}\right). \nonumber \\
\mbox{Hadamard product of} \; \bA \; \mbox{and} \; \bB &:& \bA \circ \bB = (a_{ij}\cdot b_{ij}). \nonumber \end {eqnarray}
\end {itemize}

\subsection{Basic Setup}

Suppose $\bY_1,\cdots,\bY_n$ are $p\times q$ dimensional matrix valued sample data for the treatment group, i.i.d. from a matrix normal distribution $\mathcal{MN}(\bmu_y, \bU, \bV)$, where $\bmu_y$ is the $p\times q$ dimensional mean parameter matrix, $\bU$ is a $p\times p$ dimensional covariance matrix for the column vectors of $\bY_i$, and $\bV$ is a $q\times q$ dimensional covariance matrix for the row vectors of $\bY_i$. The parameters $\bmu_y$, $\bU$ and $\bV$ are all unknown in practice. In contrast to the treatment group, we also have the control group, which consists of $p\times q$ dimensional matrix valued sample data $\bZ_1,\cdots, \bZ_m$ i.i.d. from a matrix normal distribution $\mathcal{MN}(\bmu_z,\bU, \bV)$. Compared with the treatment group, the main difference is the mean matrix $\bmu_z$. We assume most of the elements in $\bmu_y$ are the same as $\bmu_z$, but some of them are different. Furthermore, we do not know the location of these different elements. We aim to find out these signals in $\bmu_y$ compared with $\bmu_z$ through a multiple hypothesis testing framework. Another difference is that we allow the sample size $m$ of the control group to be different from that of the treatment group, which permits more flexibility in practice. 

For a generic matrix $\bA$, we use $\bA_{i,j}$ to denote the $(i,j)$th element in $\bA$. We want to simultaneously test:
\begin{equation}\label{eq4}
H_{0,ij}: \bmu_{y, ij}-\bmu_{z, ij}=0\quad\text{vs}\quad H_{1,ij}: \bmu_{y, ij}-\bmu_{z, ij}\neq 0
\end{equation}
for $i=1,\cdots, p$ and $j=1\cdots, q$.

If we focus on a single hypothesis test, suppose the standard deviation of $\bY_{l, ij}$ or $\bZ_{l, ij}$ is $\sigma_{ij}$, then for the two sample comparison, we have 
\begin{equation*}
\widetilde{\bX}_{ij}\equiv\sqrt{\frac{nm}{n+m}}(\frac{1}{n}\sum_{l=1}^n\bY_{l, ij}-\frac{1}{m}\sum_{k=1}^m\bZ_{l, ij})/\sigma_{ij}\sim\mathcal{N}(\sqrt{\frac{nm}{n+m}}(\bmu_{y,ij}-\bmu_{z,ij})/\sigma_{ij}, 1).
\end{equation*}
Therefore, in the ideal situation that $\sigma_{ij}$ is known, we can consider $\widetilde{\bX}_{ij}$ as the test statistics for hypothesis testing, and calculate p-values as $2\Phi(-|\widetilde{\bX}_{ij}|)$. However, in practice $\sigma_{ij}$ is unknown. Correspondingly, we can use pooled estimator constructed based on the two groups. More specifically, denote the sample mean $\overline{\bY}_{ij}=n^{-1}\sum_{l=1}^n\bY_{l, ij}$, and $\overline{\bZ}_{ij}=m^{-1}\sum_{k=1}^m\bZ_{k, ij}$, then 
\begin{equation}\label{eq1}
\widehat{\sigma}_{ij}^2=\frac{1}{n+m-2}\Big\{\sum_{l=1}^n(\bY_{l, ij}-\overline{\bY}_{ij})^2+\sum_{k=1}^m(\bZ_{k, ij}-\overline{\bZ}_{ij})^2\Big\}. 
\end{equation}

When we look at the whole matrix data, denote $\bSigma=(\sigma_{ij}^{-1})$ as a matrix with the $(i,j)$th element as $\sigma_{i,j}^{-1}$, then 
\begin{equation}\label{eq2}
\widetilde{\bX}\equiv\sqrt{\frac{nm}{n+m}}(\frac{1}{n}\sum_{l=1}^n\bY_l-\frac{1}{m}\sum_{k=1}^m\bZ_k)\circ\bSigma\sim\mathcal{MN}(\sqrt{\frac{nm}{n+m}}(\bmu_y-\bmu_z)\circ\bSigma, \bSigma_1,\bSigma_2),
\end{equation}
where $\bSigma_1$ and $\bSigma_2$ are the correlation matrices of $\bU$ and $\bV$, respectively. In (\ref{eq2}), the notation ``$\circ$" is Hadamard product, which means element wise product for the matrices. For the unknown marginal variances $\sigma_{ij}$, we denote $\widehat{\bSigma}$ as a matrix with the $(i,j)$th element as $\widehat{\sigma}_{ij}^{-1}$,  where each $\widehat{\sigma}_{ij}$ is defined in (\ref{eq1}). We will consider the $p\times q$ dimensional matrix 
\begin{equation*}
\bX\equiv\sqrt{\frac{nm}{n+m}}(\frac{1}{n}\sum_{l=1}^n\bY_l-\frac{1}{m}\sum_{k=1}^m\bZ_k)\circ\widehat{\bSigma}
\end{equation*}  
for the test statistics in this paper. 

In our problem, the p-value for the $(i,j)$th hypothesis is $P_{ij}=2\mathcal{F}(-|\bX_{ij}|)$, where $\mathcal{F}(\cdot)$ denotes the cumulative distribution function of the random variable $X_{ij}$. We use threshold $t$ to reject the hypotheses which have p-values smaller than $t$. Define $R(t)=\#\{P_{ij}: P_{ij}\leq t\}$ as the total number of discoveries and $V(t)=\#\{\text{true null}: P_{ij}\leq t\}$ as the number of false discoveries. Our interest focuses on approximating the false discovery proportion $\FDP(t)=V(t)/R(t)$ given an experiment, where the convention $0/0=0$ is always used in this paper. Note that given an experiment, $R(t)$ is observable, and $V(t)$ is realized but unobservable in practice. 

The challenge in this problem results from the dependence among the test statistics. Note that for the original matrix-valued sample data, the column vectors have covariance dependence denoted as $\bU$, and row vectors also possesses covariance dependence denoted as $\bV$. Furthermore, in constructing the test statistics $\bX_{ij}$, $\widehat{\sigma}_{ij}$ for $i=1,\cdots,p$, $j=1,\cdots,q$ in expression (\ref{eq1}),  are also dependent of each other. In the following sections, we will introduce two methods for approximating the $\FDP$. 

\subsection{Noodle Method}

If we vectorize $\bX$ by stacking the column vectors denoted as $vec(\bX)$, then 
\begin{equation*}
vec(\bX)=\bT^{1/2}vec(\widetilde{\bX}),
\end{equation*}
where $\bT^{1/2}$ is a $(pq)\times(pq)$ dimensional diagonal matrix. The diagonal elements of $\bT^{1/2}$ are a vectorized matrix with the $(i,j)$th element as $\sigma_{ij}/\widehat{\sigma}_{ij}$. By the property of matrix normal distribution,  based on expression (\ref{eq2}), we have 
\begin{equation}\label{xtilde}
vec(\widetilde{\bX})\sim\mathcal{N}(\sqrt{\frac{nm}{n+m}}vec((\bmu_y-\bmu_z)\circ\bSigma), \bSigma_2\otimes\bSigma_1), 
\end{equation}
where $vec(\widetilde{\bX})$ is a $pq$ dimensional column vector, the notation ``$\otimes$" denotes the Kronecker product, and $\bSigma_2\otimes\bSigma_1$ is a $(pq)\times (pq)$ dimensional covariance matrix. 

For the $vec(\widetilde{\bX})$, since it follows a multivariate normal distribution, it can be connected with a factor model structure where the random errors are weakly dependent. More specifically, applying eigenvalue decomposition to $\bSigma_2\otimes\bSigma_1$, let $\theta_1,\cdots,\theta_{pq}$ be the non-increasing eigenvalues of $\bSigma_2\otimes\bSigma_1$, and $\brho_1,\cdots,\brho_{pq}$ be the corresponding eigenvectors. We further define $\bF=(\sqrt{\theta}_1\brho_1,\cdots,\sqrt{\theta}_h\brho_h)$ for some appropriate positive integer value $h$, then $vec(\bX)$ can be expressed as
\begin{equation}\label{app1}
vec(\bX)=\bT^{1/2}(vec(\bmu^{\star})+\bF\bW+\bepsilon), 
\end{equation}
where $\bW\sim\mathcal{N}(0, \bI_h)$, $\bmu^{\star}=\sqrt{\frac{nm}{n+m}}vec((\bmu_y-\bmu_z)\circ\bSigma)$ for simplification, and $\bepsilon\sim\mathcal{N}(0,\sum_{i=h+1}^p\theta_i\brho_i\brho_i^T)$. As shown in Fan, Han \& Gu (2012), when $h$ satisfies some regularity condition, $\bepsilon$ are weakly dependent.  Note that, conditional on $\bT^{1/2}$ and $\bW$, $vec(\bX)$ are weakly dependent due to the covariance structure in $\bepsilon$. Therefore, we expect that the proportion of falsely rejected hypothesis among all tests can be approximated by $(pq)^{-1}\sum_{i\in\{\text{true null}\}}P(P_{ij}\leq t|\bT, \bW)$. Denote the diagonal elements of $\bT^{1/2}$ as $\{\sqrt{T_i}\}_{i=1}^{pq}$. By plugging the definition of p-values and noting that $T_i$ concentrates on 1 with $var(T_i)\rightarrow0$ as $n\rightarrow\infty$, we propose an approximation formula for $\FDP(t)$ by 
\begin{equation*}
\FDP_{\text{oracle,1}}(t)=\frac{1}{R(t)}\sum_{l\in\text{\{true null\}}}[\Phi(a_l(z_{t/2}+\zeta_l))+\Phi(a_l(z_{t/2}-\zeta_l))],
\end{equation*}
where $a_l=(1-\|\bff_l\|^2)^{-1/2}$, $\zeta_l=\bff_l^T\bW$ and $\bff_l^T$ is the $l$th row of $\bF$. The following Proposition 1 shows that $\FDP_{oracle,1}(t)$ is a good approximation to the true $\FDP(t)$. 

\begin{proposition}
If $(pq)^{-1}\sqrt{\theta_{h+1}^2+\cdots+\theta_{pq}^2}=O((pq)^{-\delta})$ for some $\delta>0$, $R(t)^{-1}=O_p((pq)^{-(1-\zeta)})$ for some $\zeta\geq 0$, then $|\FDP_{\text{oracle},1}(t)-\FDP(t)|=O_p((pq)^{\zeta}((pq)^{-\delta/2}+(n+m)^{-1/2})$. 
\end{proposition}
When the number of tests increases, the number of total rejections tends to increase. Here, we allow some flexibility in the growth rate through the parameter $\zeta$, so that $R(t)$ does not need to grow in the order of $pq$. However, the value of $\zeta$ should not be too large, as it will reduce the convergence rate in the FDP approximation. 

Since we do not know which hypotheses are true nulls, $\FDP_{\text{oracle},1}(t)$ can be approximated by 
\begin{equation*}
\FDP_{A,1}(t)=\frac{1}{R(t)}\sum_{l=1}^{pq}[\Phi(a_l(z_{t/2}+\zeta_l))+\Phi(a_l(z_{t/2}-\zeta_l))]. 
\end{equation*}
Here $\FDP_{A,1}(t)$ is an upper bound of $\FDP_{\text{oracle},1}(t)$. When we assume sparse signals, these two quantities will be close. 

Applying eigenvalue decomposition directly to a $(pq)\times(pq)$ dimensional matrix will be challenging. Fortunately, due to the properties of Kronecker product, the eigenvalues and eigenvectors of $\bSigma_2\otimes\bSigma_1$ in (\ref{eq2}) can be constructed based on those of $\bSigma_2$ and $\bSigma_1$. Let $\lambda_1,\cdots,\lambda_p$ be the non-increasing eigenvalues of $\bSigma_1$, and $\bnu_1,\cdots,\bnu_p$ be the corresponding eigenvectors. Let $\xi_1,\cdots,\xi_q$ be the non-increasing eigenvalues of $\bSigma_2$ and $\bgamma_1,\cdots,\bgamma_q$ be the corresponding eigenvectors. Then the eigenvalues of $\bSigma_2\otimes\bSigma_1$ are $\xi_j\times\lambda_i$ for $1\leq i\leq p$ and $1\leq j\leq q$, and the corresponding eigenvectors are $\bgamma_j\otimes\bnu_i$. 

However, in practice, the correlation matrices $\bSigma_1$ and $\bSigma_2$ in (\ref{eq2}) are both unknown. We will estimate $\bSigma_1$ by $\widehat{\bSigma}_1$:
\begin{equation*}
\widehat{\bSigma}_1=(n+m-2)^{-1}q^{-1}\{\sum_{l=1}^n((\bY_l-\overline{\bY})\circ\widehat{\bSigma})((\bY_l-\overline{\bY})\circ\widehat{\bSigma})^T+\sum_{k=1}^m((\bZ_k-\overline{\bZ})\circ\widehat{\bSigma})((\bZ_k-\overline{\bZ})\circ\widehat{\bSigma})^T\}, 
\end{equation*}
and estimate $\bSigma_2$ by $\widehat{\bSigma}_2$:
\begin{equation*}
\widehat{\bSigma}_2=(n+m-2)^{-1}p^{-1}\{\sum_{l=1}^n((\bY_l-\overline{\bY})\circ\widehat{\bSigma})^T((\bY_l-\overline{\bY})\circ\widehat{\bSigma})+\sum_{k=1}^n((\bZ_k-\overline{\bZ})\circ\widehat{\bSigma})^T((\bZ_k-\overline{\bZ})\circ\widehat{\bSigma})\},
\end{equation*}
where $\overline{\bY}=n^{-1}\sum_{l=1}^n\bY_l$ and $\overline{\bZ}=m^{-1}\sum_{k=1}^m\bZ_k$. These are pooled sample correlation estimators. For diverging $p$ and $q$, $\widehat{\bSigma}_1$ and $\widehat{\bSigma}_2$ are not necessarily consistent estimates of $\bSigma_1$ and $\bSigma_2$, respectively. However, we will show that for FDP approximation, $\widehat{\bSigma}_1$ and $\widehat{\bSigma}_2$ can still lead to good approximation results. 

Let $\widehat{\lambda}_1,\cdots,\widehat{\lambda}_p$ be the eigenvalues of $\widehat{\bSigma}_1$, and $\widehat{\bnu}_1,\cdots,\widehat{\bnu}_p$ be the corresponding eigenvectors. Let $\widehat{\xi}_1,\cdots,\widehat{\xi}_q$ be the eigenvalues of $\widehat{\bSigma}_2$, and $\widehat{\bgamma}_1,\cdots,\widehat{\bgamma}_q$ be the corresponding eigenvectors. For the eigenvalues $\{\widehat{\lambda}_i\}$ and $\{\widehat{\xi}_j\}$, we calculate the product of each possible pair to obtain the eigenvalues of $\widehat{\bSigma}_2\otimes\widehat{\bSigma}_1$, and arrange these values in a non-increasing order, written as $\{\widehat{\theta}_l\}$. Correspondingly, the eigenvectors of $\widehat{\bSigma}_2\otimes\widehat{\bSigma}_1$ will be written as $\{\widehat{\brho}_l\}$. For a given integer value $h$, we define $(pq)\times h$ dimensional matrix $\widehat{\bF}=(\sqrt{\widehat{\theta}_1}\widehat{\brho}_1,\cdots,\sqrt{\widehat{\theta}_h}\widehat{\brho}_h)$. Given an experiment, $\bW$ is a realized but unobserved vector. We will consider a least squares estimator of $\bW$: $\widehat{\bW}=(\widehat{\bF}^T\widehat{\bF})^{-1}\widehat{\bF}^Tvec(\bX)$. Then we can approximate the $\FDP_{A,1}(t)$ by 
\begin{equation*}
\widehat{\FDP}_1(t)=\frac{1}{R(t)}\sum_{l=1}^{p\times q}[\Phi(\widehat{a}_l(z_{t/2}+\widehat{\zeta}_l))+\Phi(\widehat{a}_l(z_{t/2}-\widehat{\zeta}_l))], 
\end{equation*}
where $\widehat{a}_l=(1-\|\widehat{\bff}_l\|^2)^{-1/2}$, $\widehat{\zeta}_l=\widehat{\bff}_l^T(\widehat{\bF}^T\widehat{\bF})^{-1}\widehat{\bF}^Tvec(\bX)$ and $\widehat{\bff}_l^T$ is the $l$th row of $\widehat{\bF}$. 

\begin{theorem}
Under the conditions in Proposition 1, in addition, $\theta_i-\theta_{i+1}\geq g_{pq}$ with positive $g_{pq}\asymp pq$ for $i=1,\cdots, h$, $\{\widehat{a}_l\}_{l=1}^{pq}$ and $\{a_l\}_{l=1}^{pq}$ are upper bounded, then $|\widehat{\FDP}_1(t)-\FDP_{A,1}(t)|=O_p((pq)^{\zeta}(h(n+m)^{-1/2}+(pq)^{-1/2}\|vec(\bmu^{\star})\|))$. 
\end{theorem}

In Theorem 1, we require an eigengap condition for the largest $h$ eigenvalues. Fan, Liao and Mincheva (2013) has shown that such condition can be satisfied for factor model structures. In the FDP approximation, the convergence rate also depends on the magnitude of signals, $\|vec(\bmu^{\star})\|$. When we consider sparse signals in the mean matrix for the two group comparison, we expect that $(pq)^{-1/2}\|vec(\bmu^{\star})\|$ converges to zero. 

To determine $h$, we can use the eigenvalue ratio estimator in Ahn \& Horenstein (2013). The estimator is $\widehat{h}=\argmax_{1\leq l\leq l_{\max}}(\widehat{\theta}_l/\widehat{\theta}_{l+1})$, where $l_{\max}$ is a pre-determined maximum possible number of factors. Ahn \& Horenstein (2013) has shown that under mild regularity conditions, the eigenvalue ratio estimator is consistent for the true number of factors.

In practice, if we have a priori knowledge for the two correlation matrices, $\bSigma_1$ and $\bSigma_2$, the regularity condition in Theorem 1 can be substantially relaxed. For example, if we know that $\bSigma_1$ and $\bSigma_2$ are sparse matrices (Bickel and Levina 2008), we can propose consistent thresholding estimators for $\bSigma_1$ and $\bSigma_2$. Correspondingly, the eigengap condition in Theorem 1 can be relaxed to $d_{pq}\asymp d$ for $i=1,\cdots, h$ where $d$ is a constant. Under such scenario, we can still achieve the FDP approximation results in Theorem 1. In some extreme cases, e.g., $\bSigma_1$ and $\bSigma_2$ are both identity matrices, each element in the matrix valued data is independent of each other. Then in Proposition 1, we can choose $h=0$. Correspondingly, the condition $(pq)^{-1}\sqrt{\theta_{h+1}^2+\cdots+\theta_{pq}^2}=(pq)^{-1/2}$. Since we have the priori knowledge that $\bSigma_1$ and $\bSigma_2$ are identity matrices, we do not need to estimate them by $\widehat{\bSigma}_1$ and $\widehat{\bSigma}_2$. The FDP approximation will be simplified as $\widehat{\FDP}(t)=pqt/R(t)$. If we further replace $pq$ by an estimate of the number of true signals, it can be connected with the Storey's procedure (Storey 2002). Nevertheless, the FDP approximation that we proposed here is more general, especially designed for strong dependence scenarios. 

We call the above described procedure as noodle method, as we cut the ``dough" (matrix valued data) into slices (column vectors) and stick into a long ``noodle". 

\subsection{Sandwich Method} 

In the noodle method, the approximation of $\FDP$ relies on the eigenvalues and eigenvectors of $\widehat{\bSigma}_2\otimes\widehat{\bSigma}_1$. Note that we need to calculate the Kronecker product of each possible pair from $\{\widehat{\bgamma}_j\}$ and $\{\widehat{\bnu}_i\}$. When $pq$ is large, this step can be very computationally intensive. The question is whether we can provide an alternative approximation procedure for $\FDP$ which is more computationally efficient. The key idea in principal factor approximation in Fan, Han and Gu (2012) is to use the first few principal components to capture the majority dependence among the test statistics. In our paper, the problem is somewhat different, because there are two covariance matrices, $\bSigma_1$ and $\bSigma_2$ for modeling the column and row dependence, respectively. If we can use the first few principal components from $\widehat{\bSigma}_1$ and from $\widehat{\bSigma}_2$ respectively to capture the majority dependence among the test statistics, the computation will be substantially simplified, and the corresponding procedure will be very appealing. This motivates us to propose the following method. 

Note that in expression (\ref{eq2}), $\widetilde{\bX}\sim\mathcal{MN}(\bmu^{\star},\bSigma_1,\bSigma_2)$ if we let $\bmu^{\star}=\sqrt{\frac{nm}{n+m}}vec((\bmu_y-\bmu_z)\circ\bSigma)$ to simplify the notation. Then for some positive integer values $k_1$ and $k_2$, we can rewrite $\widetilde{\bX}$ as 
\begin{equation}\label{eq5}
\widetilde{\bX}=\bmu^{\star}+\bC\widetilde{\bW}\bD+\bepsilon
\end{equation}
where $\bC=(\sqrt{\lambda_1}\bnu_1,\cdots,\sqrt{\lambda_{k_1}}\bnu_{k_1})$, $\bD=(\sqrt{\xi_1}\bgamma_1,\cdots,\sqrt{\xi_{k_2}}\bgamma_{k_2})^T$, $\widetilde{\bW}\sim\mathcal{MN}(\bzero,\bI_{k_1},\bI_{k_2})$, and $\bepsilon\sim\mathcal{MN}(\bzero,\sum_{i=k_1+1}^p\lambda_i\bnu_i\bnu_i^T,\sum_{j=k_2+1}^q\xi_j\bgamma_j\bgamma_j^T)$. 

Note that $\bC$ contains the first $k_1$ principal components from $\bSigma_1$, and $\bD$ contains the first $k_2$ principal components from $\bSigma_2$. Compared with the factor model structure in Fan, Han and Gu (2012), we have $\bC$ and $\bD$ here to capture the column and the row dependences, respectively. When $k_1$ and $k_2$ are appropriately chosen, the covariance matrices in $\bepsilon$ are both weakly dependent. 

By the properties of Kronecker product, vectorizing expression (\ref{eq5}) leads to 
\begin{equation}\label{vecx}
vec(\widetilde{\bX})=vec(\bmu)+(\bD^T\otimes\bC)vec(\widetilde{\bW})+vec(\bepsilon). 
\end{equation}
Similar to the discussion in section 2.3, we can propose an approximation for $\FDP(t)$ as 
\begin{equation*}
\FDP_{\text{oracle},2}(t)=\frac{1}{R(t)}\sum_{l\in\{\text{true null}\}}[\Phi(d_l(z_{t/2}+\eta_l))+\Phi(a_l(z_{t/2}-\eta_l))]
\end{equation*}
where $\eta_l=\bb_l^Tvec(\widetilde{\bW})$, $\bb_l$ is the $l$th row of $\bD^T\otimes\bC$, and $d_l=(1-\|\bb_l\|^2)^{-1/2}$. The following Proposition 2 shows that $\FDP_{\text{oracle},2}$ is also a good approximation for the true $\FDP$. 

\begin{proposition}
If $p^{-1}\sqrt{\lambda_{k_1+1}^2+\cdots+\lambda_p^2}=O(p^{-\delta_1})$ and $q^{-1}\sqrt{\xi_{k_2+1}^2+\cdots+\xi_q^2}=O(q^{-\delta_2})$ for some $\delta_1>0$ and $\delta_2>0$, $R(t)^{-1}=O_p((pq)^{-(1-\zeta)})$ for some $\zeta\geq 0$, then $|\FDP_{\text{oracle},2}(t)-\FDP(t)|=O_p((pq)^{\zeta}(p^{-\delta_1/2}q^{-\delta_2/2}+(n+m)^{-1/2}))$. 
\end{proposition}

Replacing the summation over true nulls in $\FDP_{\text{oracle},2}(t)$ by all the tests, we have an upper bound as 
\begin{equation*}
\FDP_{A,2}(t)=\frac{1}{R(t)}\sum_{l=1}^{pq}[\Phi(d_l(z_{t/2}+\zeta_l))+\Phi(d_l(z_{t/2}-\zeta_l))]. 
\end{equation*}

If $\bSigma_1$ and $\bSigma_2$ are known, we can estimate the realized $vec(\widetilde{\bW})$ by least squares estimator
\begin{equation*}
[(\bD^T\otimes\bC)^T(\bD^T\otimes\bC)]^{-1}(\bD^T\otimes\bC)^Tvec(\bX). 
\end{equation*}

For unknown $\bSigma_1$ and $\bSigma_2$, we use $\widehat{\bSigma}_1$ and $\widehat{\bSigma}_2$ for the estimation. Correspondingly, we replace $\bC$ and $\bD$ by $\widehat{\bC}$ and $\widehat{\bD}$ respectively, where eigenvalues and eigenvectors are replaced by their estimates. Furthermore, we consider the $\FDP$ approximation formula:
\begin{equation*}
\widehat{\FDP}_2(t)=\sum_{l=1}^{pq}[\Phi(\widehat{d}_l(z_{t/2}+\widehat{\eta}_l))+\Phi(\widehat{d}_l(z_{t/2}-\widehat{\eta}_l))]/R(t)
\end{equation*}
where $\widehat{d}_l=(1-\|\widehat{\bb}_l\|^2)^{-1/2}$, $\widehat{\bb}_l$ is the $l$th row of $\widehat{\bD}^T\otimes\widehat{\bC}$, and $\widehat{\eta}_l$ is the $l$th element of $[(\sum_{i=1}^{k_2}\widehat{\bgamma}_i\widehat{\bgamma}_i^T)\otimes(\sum_{j=1}^{k_1}\widehat{\bnu}_j\widehat{\bnu}_j^T)]vec(\bX)$. 

It is worth mentioning in $\widehat{\FDP}_2(t)$, we only need to calculate the Kronecker product of the first few eigenvectors from $\widehat{\bSigma}_1$ and $\widehat{\bSigma}_2$. This can avoid the computational issue in noodle method for large values of $p$ and $q$, where Kronecker product has to be calculated for all possible pairs. 

\begin{theorem}
Under the conditions in Proposition 2, in addition, $\lambda_i-\lambda_{i+1}\geq g_p$ with $g_p\asymp p$ for $i=1,\cdots, k_1$ and $\xi_j-\xi_{j+1}\geq g_q$ with $g_q\asymp q$ for $j=1,\cdots, k_2$, $\{\widehat{d}\}_{l=1}^{pq}$ and $\{d_l\}_{l=1}^{pq}$ are upper bounded, then $|\widehat{\FDP}_2(t)-\FDP_{A,2}(t)|=O_p((pq)^{\eta}(k_1k_2(n+m)^{-1}+(k_1+k_2)(n+m)^{-1/2}+(pq)^{-1/2}\|vec(\bmu^{\star})\|)$. 
\end{theorem}

To determine $k_1$ and $k_2$, we consider the eigenvalue ratio estimator:
\begin{equation*}
\widehat{k}_1=\argmax_{1\leq l\leq l_{\max}}(\widehat{\lambda}_l/\widehat{\lambda}_{l+1}), \quad\quad\widehat{k}_2=\argmax_{1\leq l\leq l_{\max}}(\widehat{\xi}_l/\widehat{\xi}_{l+1}).
\end{equation*}

As discussed in section 2.3, if we have a priori knowledge for the two correlation matrices, the eigengap condition in Theorem 2 can be relaxed. The key idea in the above procedure is to express the matrix data in terms of a ``sandwich" formula (\ref{eq5}), where the two matrices $\bC$ and $\bD$ ``wrap" the common factor matrix $\bW$. Thus, we call the proposed procedure in section 2.3 sandwich method. 

\subsection{Trimmed Regression}

In noodle method and sandwich method, we consider the least squares estimator for the realized $\bW$ and $\widetilde{\bW}$. Correspondingly, the convergence rate in the FDP approximation involves the term $\|vec(\bmu^{\star})\|$. In practice, we may consider a trimmed method to further reduce the effect from $\bmu^{\star}$. According to (\ref{app1}), we will sort the absolute values of $vec(\bX)$ in a non-decreasing order, and denote them as $Z_1, \cdots, Z_{pq}$. For the nonzero elements in $vec(\bmu^{\star})$, the corresponding $Z_i$'s tend to be larger. We can choose the smallest $m$ $Z_i$'s. To simplify the notation, approximately we can write the unified expression
\begin{equation*}
Z_i=\bb_i^T\bW^{\star}+\epsilon_i, \quad\quad i=1,\cdots,m
\end{equation*}
where $\bb_i^T$'s are the corresponding rows in $\bB$.  The matrix $\bB$ refers to $\bF$ in the noodle method and $\bD^T\otimes\bC$ in the sandwich method. The $\bW$ refers to $\bW$ in the noodle method and $vec(\widetilde{\bW})$ in the sandwich method. The unknown $\bW$ will be estimated based on the following robust $L_1$ regression:
\begin{equation*}
\widehat{\bW}^{\star}=\argmin_{W}\frac{1}{m}\sum_{l=1}^m\Big|Z_l-\widehat{\bb}_l^T\bW\Big|, 
\end{equation*}
where $\widehat{\bb}_l^T$'s are the rows of $\widehat{\bB}$. The above method has been incorporated to our proposed methods in the following numerical studies. 

\section{Simulation Studies}
In our simulation studies,      
The treatment group data are generated from $\mathbf{Y}_i \sim \mathcal{MN}(\bmu,\mathbf{\Sigma_1},\mathbf{\Sigma_2}), i=1,...,n$ and the control group data are generated from $\mathbf{Z}_j \sim \mathcal{MN}(\bzero,\mathbf{\Sigma_1},\mathbf{\Sigma_2}), j=1,...,m$. The signal strength $\bmu$ equals 1 for the first 8 rows out of p rows, first 25 columns out of q columns, and 0 otherwise. We consider sample size n=m=50, dimensionality p=q=100 for both noodle method and sandwich method unless stated otherwise, threshold value t=0.001 and the number of simulation rounds to be 500. We estimate the unknown number of factors by the data-driven eigenvalue ratio method for noodle method and sandwich method, both with $k_{max} = \lfloor 0.2(n+m)\rfloor$. We now examine the performance of our two methods on simulated data sets, which are constructed under the framework of three models. 
   
\noindent{\bf Model 1:}
   
Let $\mathbf{B}_1$ be a $p \times l_1$ dimensional matrix with each element generated from a distribution $\mathcal{F}_1$, $\Sigma_{u1}$ be a $p \times p$ dimensional diagonal matrix with all diagonal values being 0.5, then $\mathbf{\Sigma}_1$ is the correlation matrix of $\mathbf{B}_1\mathbf{B}_1^T + \Sigma_{u1}$; Similarly, let $\mathbf{B}_2$ be a $q \times l_2$ dimensional matrix with each element generated from a distribution $\mathcal{F}_2$, $\Sigma_{u2}$ be a $q \times q$ dimensional diagonal matrix with all diagonal values being 0.5, then $\mathbf{\Sigma}_2$ is the correlation matrix of $\mathbf{B}_2\mathbf{B}_2^T + \Sigma_{u2}$
\begin{itemize}
\item $l_1=l_2=3$, $\mathcal{F}_1$ and $\mathcal{F}_2$ are both $\mathcal{N}(0,1)$. 
\item $l_1=2, l_2=4$, $\mathcal{F}_1$ and $\mathcal{F}_2$ are both $Uniform(0,1)$. 
\end{itemize}

In Model 1, both $\bSigma_1$ and $\bSigma_2$  possess some strict factor model structures. 
  
\noindent{\bf Model 2:}

We keep the similar setting in Model 1, but consider $\bSigma_{u1}$ to be a $p \times p$ dimensional power decay matrix with $\rho_1$, where the $(i,j)$th element of $\bSigma_{u1}$ is defined as $\rho_1^{|i-j|}$. Similarly, let $\bSigma_{u2}$ be a $q \times q$ dimensional power decay matrix with $\rho_2$.  In Model 2, we consider $l_1=l_2=3$, and $\mathcal{F}_1$, $\mathcal{F}_2$ are both $Uniform(0,1)$. We examine two settings: $(\rho_1,\rho_2)=(0.5,0.3)$ and $(\rho_1,\rho_2)=(0.5,0.8)$.

In Model 2, both $\bSigma_1$ and $\bSigma_2$ possess some approximate factor model structures, which can be used for testing the robustness of the eigenvalue ratio estimator for the unknown number of factors under the matrix normal settings. 

\noindent{\bf Model 3:}

We keep the similar setting in model 1, where both $\mathcal{F}_1$ and $\mathcal{F}_2$ are $Uniform(0,1)$. Let $(\lambda_1,...,\lambda_p)$ and $(\nu_1,...\nu_p)$ be the eigenvalues and the corresponding eigenvectors of $\mathbf{\Sigma}_1$, respectively.  Let $(\xi_1,...,\xi_q)$ and $(\gamma_1,...\gamma_q)$ be the eigenvalues and the corresponding eigenvectors of $\mathbf{\Sigma_2}$, respectively. Define $\widetilde{\mathbf{C}}=(\sqrt{\lambda_1}\nu_1,...,\sqrt{\lambda_p}\nu_p)$, and $\widetilde{\mathbf{D}}=(\sqrt{\xi_1}\gamma_1,...,\sqrt{\xi_q}\gamma_q)$. Then the data matrix $\mathbf{X}$ is generated from $\mathbf{X}=\mathbf{\mu} + \widetilde{\mathbf{C}}\mathbf{W}\widetilde{\mathbf{D}}$. In our simulation studies, we consider following settings:
\begin{itemize}
       \item $(l_1,l_2)=(2,2), (3,3), (4,4), (2,4)$ as the choices for $\mathbf{B}_1$ and $\mathbf{B}_2$;
       \item $\mathbf{W}$ is a $p \times q$ dimensional matrix with each element randomly generated from $\sqrt{\frac{2}{3}}t_6$ distribution, or exponential distribution with $\lambda=1$. 
\end{itemize}

Model 3 is designed to test the performance of our proposed methods when the matrix normality assumption is violated.    
   
We will compare our newly proposed methods with PFA method in Fan and Han (2017).  The PFA method was originally designed for vector data from multivariate normal distribution. In the current paper,  we are dealing with matrix variated data $\mathbf{Y}$ and $\mathbf{Z}$.  To apply the PFA method, we vectorize $\mathbf{Y}$ and $\mathbf{Z}$ to obtain $vec(\mathbf{Y})$ and $vec(\mathbf{Z})$, and assume $vec(\bY)\sim \mathcal{N}(vec(\bmu),\bSigma)$, $vec(\bZ)\sim \mathcal{N}(\bzero,\bSigma)$. We need to estimate the $(pq) \times (pq)$ dimensional covariance matrix $\bSigma$ based on the sample data $\{vec(\mathbf{Y}_i)\}_{i=1}^n$ and $\{vec(\mathbf{Z}_j)\}_{j=1}^m$. Suppose we consider the pooled sample covariance matrix as our estimator, then
    \begin{eqnarray*}
        S &=& \frac{1}{n+m-2}\Big\{\sum_{i=1}^n (vec(\mathbf{Y}_i)-\overline{vec(\mathbf{Y})})(vec(\mathbf{Y}_i)-\overline{vec(\mathbf{Y})})^T\\
        &&\quad\quad\quad\quad+ \sum_{j=1}^m (vec(\mathbf{Z}_j)-\overline{vec(\mathbf{Z})})(vec(\mathbf{Z}_j)-\overline{vec(\mathbf{Z})})^T  \Big \}
    \end{eqnarray*}
where $\overline{vec(\mathbf{Y})}$ and $\overline{vec(\mathbf{Z})}$ are the sample means of $vec(\mathbf{Y}_1),...,vec(\mathbf{Y}_n)$ and $vec(\mathbf{Z}_1),...,vec(\mathbf{Z}_m)$, respectively. To apply PFA method, we need to get the eigenvalues and eigenvectors of $S$. However, since $p \times q$ is large, it will be time-consuming to directly apply eigenvalue decomposition to $S$. Instead, we consider 
    \begin{align}
        F = \sqrt{\frac{1}{n+m-2}}(\mathbf{Y}_1^{\nu},...,\mathbf{Y}_n^{\nu},\mathbf{Z}_1^{\nu},...,\mathbf{Z}_m^{\nu})
    \end{align}
    where $\mathbf{Y}_i^{\nu}=vec(\mathbf{Y}_i)-\overline{vec(\mathbf{Y})}$ and $\mathbf{Z}_j^{\nu}=vec(\mathbf{Z}_j)-\overline{vec(\mathbf{Z})}$ for $i=1,...,n, j=1,...,m$. Then clearly, $S = FF^T$. Eigenvalue decomposition of $F$ will provide the eigenvalues and eigenvectors of $S$. Such construction would reduce the computation complexity.

\begin{figure}[h!!!]
\begin{tabular}{ccc}
  \includegraphics[width=50mm]{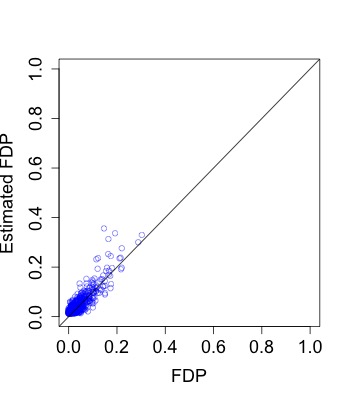} &   \includegraphics[width=50mm]{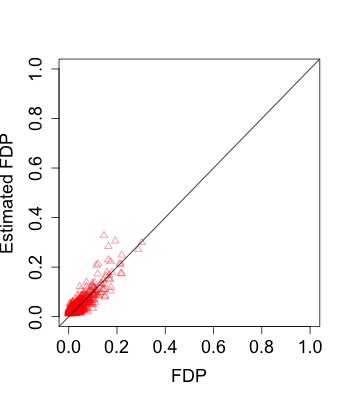} &
  \includegraphics[width=50mm]{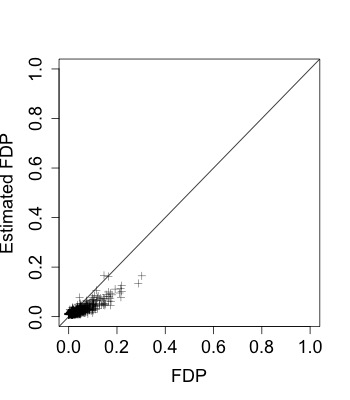}  \\
\small{(a) f(2,4), $\mathbf{B} \sim  U(-1,1)$} &
\small{(b) f(2,4), $\mathbf{B} \sim  U(-1,1)$} &
\small{(c) f(2,4), $\mathbf{B} \sim  U(-1,1)$}\\[6pt]
 \includegraphics[width=50mm]{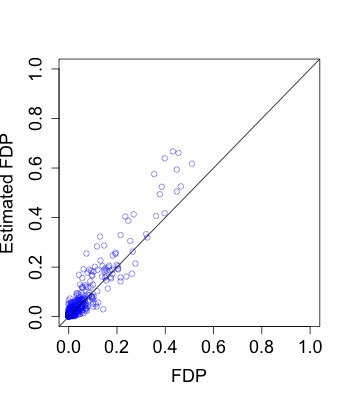} &   \includegraphics[width=50mm]{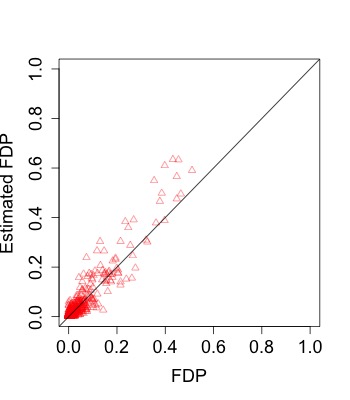} &
 \includegraphics[width=50mm]{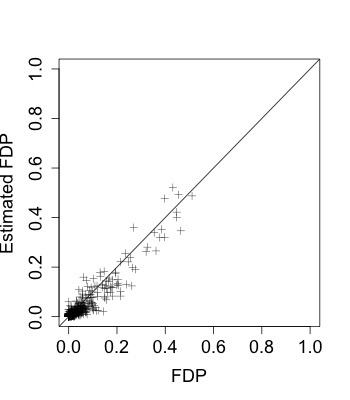}\\
\small{(d) f(3,3), $\mathbf{B} \sim N(0,1)$ } & 
\small{(e) f(3,3), $\mathbf{B} \sim N(0,1)$ } &
\small{(f) f(3,3), $\mathbf{B} \sim N(0,1)$ }\\[6pt]
\end{tabular}
\centering
\caption{\footnotesize{Model 1, the estimated values of FDP obtained by noodle method (blue circle), sandwich method (red triangle) and PFA (black crossover). The first row shows the first setting of Model 1, and the second row shows the second setting of Model 1. Here, n=m=50, p=q=100, and t=0.001.}}
\end{figure}

We compare the estimated false discovery proportion from both of our proposed methods and PFA with the true value of false discovery proportion. The results are summarized in Figures 1-6 and Table 1. In Figures 1-6, points closer to the diagonal line suggest good approximation. Under various settings, both of our proposed methods produce points slightly above the diagonal line, while the points from PFA are generally under the diagonal one. This phenomenon is further confirmed by the results in Table 1, where we calculate the mean difference between the estimated FDP and the true FDP over the 500 simulation rounds. Table 1 shows that our new estimator performs better than PFA estimator in the sense that, PFA estimator dramatically underestimates the true $\FDP$, while our new method consistently overestimates the true $\FDP$ a little bit. That means, our new estimator can provide an upper bound for estimating $\FDP$, which is meaningful in practice. It is worth mentioning that both noodle method and sandwich method perform roughly the same, but sandwich method is much more computationally efficient.\\

 
\begin{figure}[h!!!]
\begin{tabular}{ccc}
  \includegraphics[width=50mm]{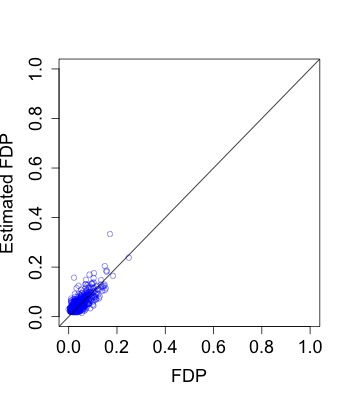} &   \includegraphics[width=50mm]{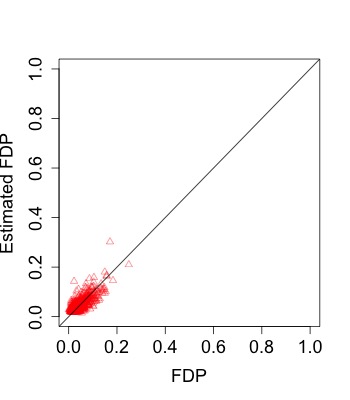} &
  \includegraphics[width=50mm]{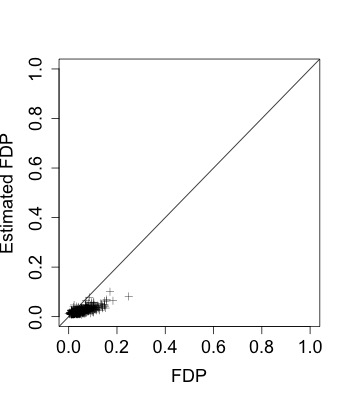}  \\
\small{(a) $(\rho_1,\rho_2) = (0.5,0.3)$} &
\small{(b) $(\rho_1,\rho_2) = (0.5,0.3)$} &
\small{(c) $(\rho_1,\rho_2) = (0.5,0.3)$}\\[6pt]
 \includegraphics[width=50mm]{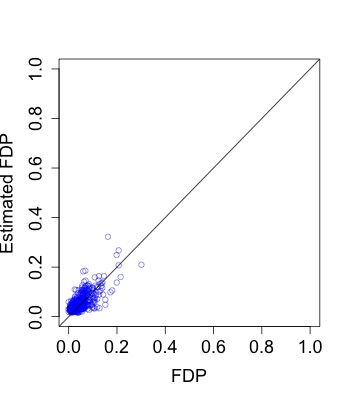} &   \includegraphics[width=50mm]{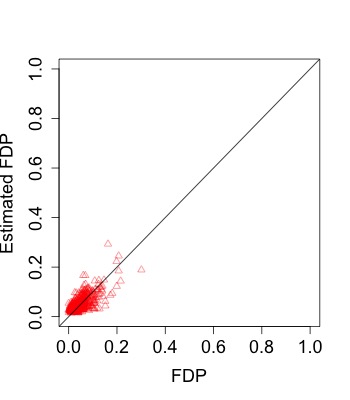} &
 \includegraphics[width=50mm]{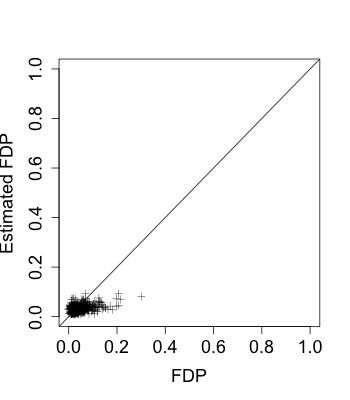}\\
\small{(d) $(\rho_1,\rho_2) = (0.5,0.8)$ } & 
\small{(e) $(\rho_1,\rho_2) = (0.5,0.8)$ } &
\small{(f) $(\rho_1,\rho_2) = (0.5,0.8)$ }\\[6pt]
\end{tabular}
\centering
\caption{\footnotesize{Model 2, the estimated values of FDP obtained by noodle method (blue circle), sandwich method (red triangle) and PFA (black crossover). The first row shows the first setting of Model 2, and the second row shows the second setting of Model 2. Here, n=m=50, p=q=100, and t=0.001.}}
\end{figure}

\begin{figure}[h!!!]
\begin{tabular}{ccc}
  \includegraphics[width=50mm]{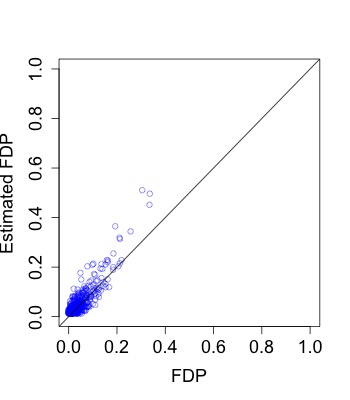} &   \includegraphics[width=50mm]{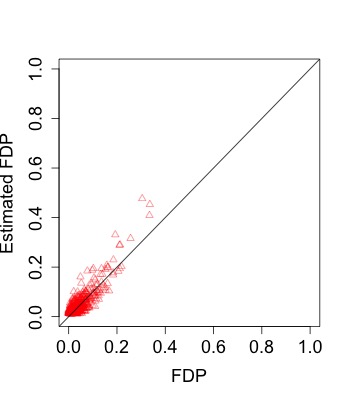} &
  \includegraphics[width=50mm]{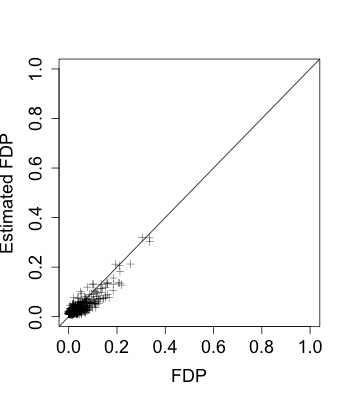} \\
\small{(a) $f(2,2), \mathbf{W} \sim Exp(1)$} &
\small{(b) $f(2,2), \mathbf{W} \sim Exp(1)$} &
\small{(c) $f(2,2), \mathbf{W} \sim Exp(1)$}
\\[6pt]
 \includegraphics[width=50mm]{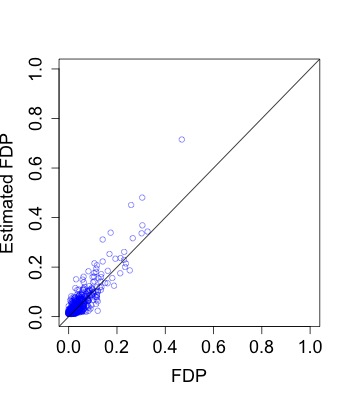} &   \includegraphics[width=50mm]{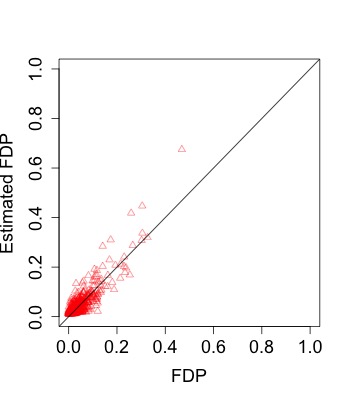} &
 \includegraphics[width=50mm]{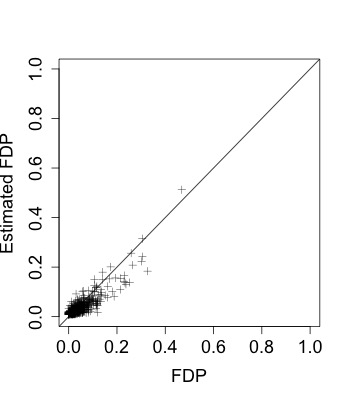}\\
\small{(d) $f(2,2), \mathbf{W} \sim \sqrt{\frac{2}{3}}t_6$ } & 
\small{(e) $f(2,2), \mathbf{W} \sim \sqrt{\frac{2}{3}}t_6$ } &
\small{(f) $f(2,2), \mathbf{W} \sim \sqrt{\frac{2}{3}}t_6$ }
\\[6pt]
\end{tabular}
\centering
\caption{\footnotesize{Model 3: factor (2,2) case, the estimated values of FDP obtained by noodle method (blue circle), sandwich method (red triangle) and PFA (black crossover). The first row represents the case where $\mathbf{W}$ with each element following Exp(1) distribution, and the second row shows the case where $\mathbf{W}$ with each element following $\sqrt{\frac{2}{3}}t_6$ distribution. Here, n=m=50, p=q=100, and t=0.001.}}
\end{figure}

\begin{figure}[h!!!]
\begin{tabular}{ccc}
  \includegraphics[width=50mm]{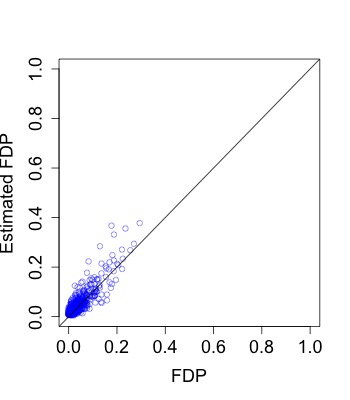} &   \includegraphics[width=50mm]{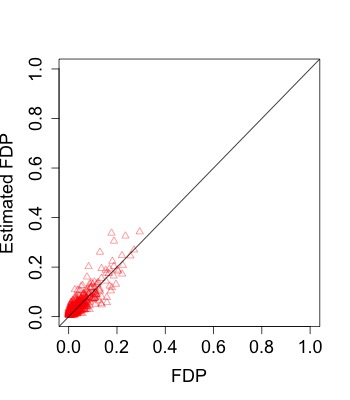} &
  \includegraphics[width=50mm]{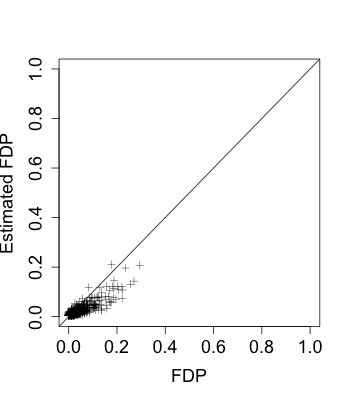} \\
\small{(a) $f(3,3), \mathbf{W} \sim Exp(1)$} &
\small{(b) $f(3,3), \mathbf{W} \sim Exp(1)$} &
\small{(c) $f(3,3), \mathbf{W} \sim Exp(1)$}
\\[6pt]
 \includegraphics[width=50mm]{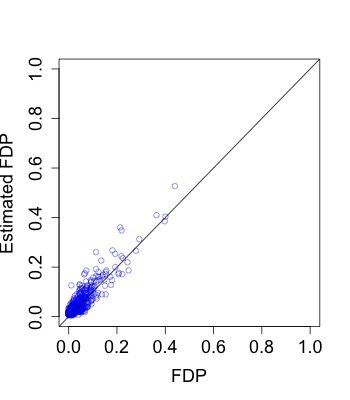} &   \includegraphics[width=50mm]{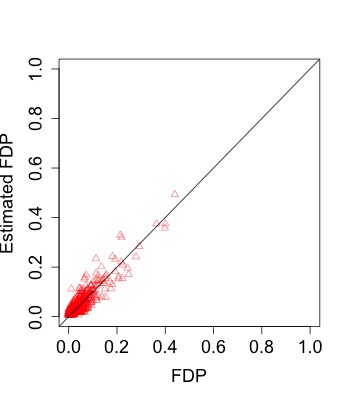} &
 \includegraphics[width=50mm]{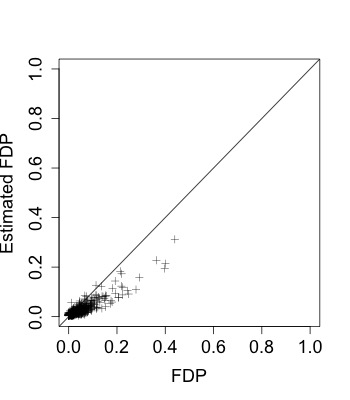}\\
\small{(d) $f(3,3), \mathbf{W} \sim \sqrt{\frac{2}{3}}t_6$ } & 
\small{(e) $f(3,3), \mathbf{W} \sim \sqrt{\frac{2}{3}}t_6$ } &
\small{(f) $f(3,3), \mathbf{W} \sim \sqrt{\frac{2}{3}}t_6$ }
\\[6pt]
\end{tabular}
\centering
\caption{\footnotesize{Model 3: factor (3,3) case, the estimated values of FDP obtained by noodle method (blue circle), sandwich method (red triangle) and PFA (black crossover). The first row represents the case where $\mathbf{W}$ with each element following Exp(1) distribution, and the second row shows the case where $\mathbf{W}$ with each element following $\sqrt{\frac{2}{3}}t_6$ distribution. Here, n=m=50, p=q=100, and t=0.001.}}
\end{figure}

\begin{figure}[h!!!]
\begin{tabular}{ccc}
  \includegraphics[width=50mm]{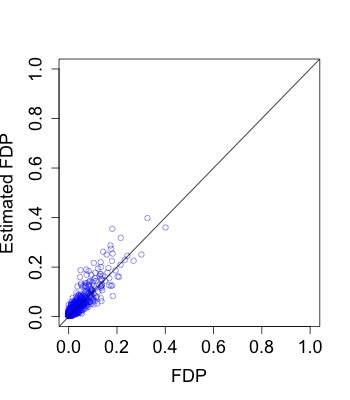} &   \includegraphics[width=50mm]{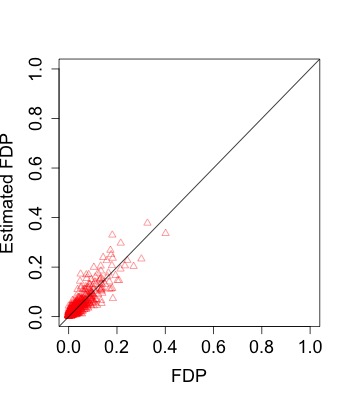} &
  \includegraphics[width=50mm]{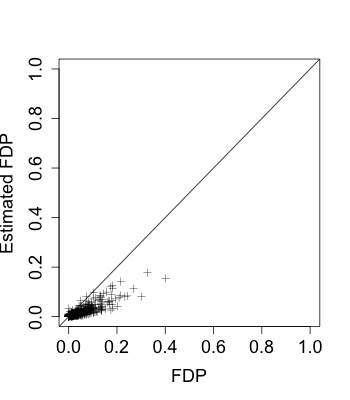}  \\
\small{(a) $f(4,4), \mathbf{W} \sim Exp(1)$} &
\small{(b) $f(4,4), \mathbf{W} \sim Exp(1)$} &
\small{(c) $f(4,4), \mathbf{W} \sim Exp(1)$}
\\
 \includegraphics[width=50mm]{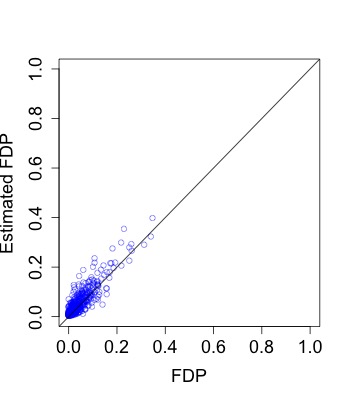} &   \includegraphics[width=50mm]{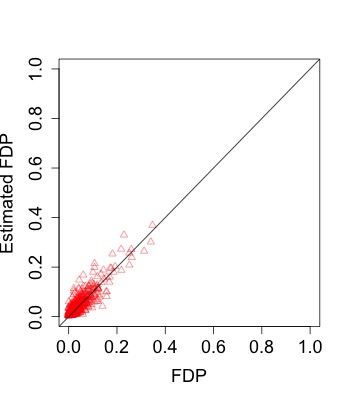} &
 \includegraphics[width=50mm]{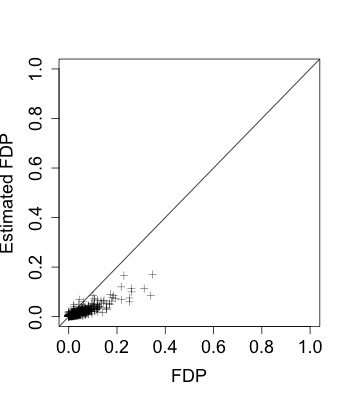}\\
\small{(d) $f(4,4), \mathbf{W} \sim \sqrt{\frac{2}{3}}t_6$ } & 
\small{(e) $f(4,4), \mathbf{W} \sim \sqrt{\frac{2}{3}}t_6$ } &
\small{(f) $f(4,4), \mathbf{W} \sim \sqrt{\frac{2}{3}}t_6$ }
\\
\end{tabular}
\centering
\caption{\footnotesize{Model 3: factor (4,4) case, the estimated values of FDP obtained by noodle method (blue circle), sandwich method (red triangle) and PFA (black crossover). The first row represents the case where $\mathbf{W}$ with each element following Exp(1) distribution, and the second row shows the case where $\mathbf{W}$ with each element following $\sqrt{\frac{2}{3}}t_6$ distribution. Here, n=m=50, p=q=100, and t=0.001.}}
\end{figure}

\begin{figure}[h!!!]
\begin{tabular}{ccc}
  \includegraphics[width=50mm]{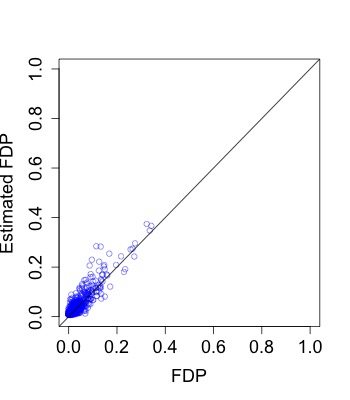} &   \includegraphics[width=50mm]{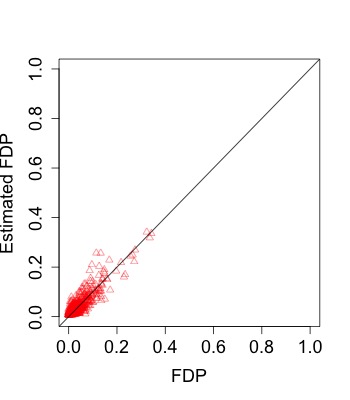} &
  \includegraphics[width=50mm]{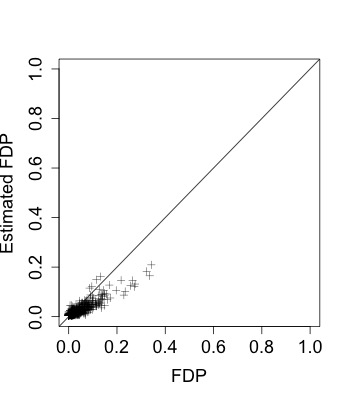}  \\
\small{(a) $f(2,4), \mathbf{W} \sim Exp(1)$} &
\small{(b) $f(2,4), \mathbf{W} \sim Exp(1)$} &
\small{(c) $f(2,4), \mathbf{W} \sim Exp(1)$}
\\
 \includegraphics[width=50mm]{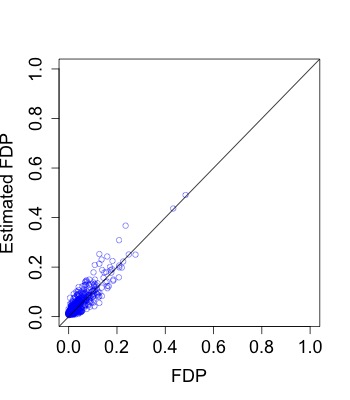} &   \includegraphics[width=50mm]{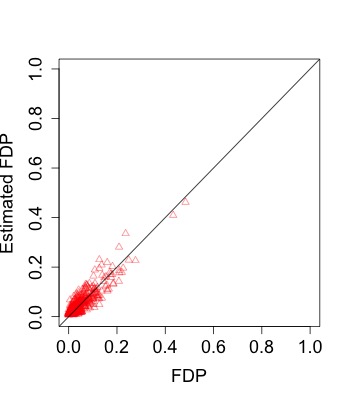} &
 \includegraphics[width=50mm]{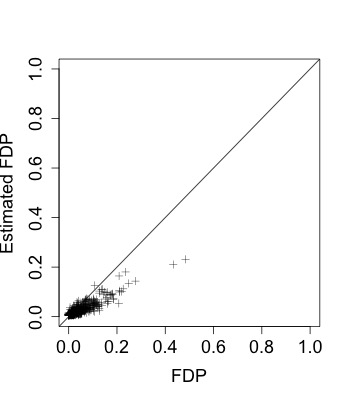}\\
\small{(d) $f(2,4), \mathbf{W} \sim \sqrt{\frac{2}{3}}t_6$ } & 
\small{(e) $f(2,4), \mathbf{W} \sim \sqrt{\frac{2}{3}}t_6$ } &
\small{(f) $f(2,4), \mathbf{W} \sim \sqrt{\frac{2}{3}}t_6$ }
\\
\end{tabular}
\centering
\caption{\footnotesize{Model 3: factor (2,4) case, the estimated values of FDP obtained by noodle method (blue circle), sandwich method (red triangle) and PFA (black crossover). The first row represents the case where $\mathbf{W}$ with each element following Exp(1) distribution, and the second row shows the case where $\mathbf{W}$ with each element following $\sqrt{\frac{2}{3}}t_6$ distribution. Here, n=m=50, p=q=100, and t=0.001.}}
\end{figure}

The above analysis is based on the setting that the dimensionality $p$ and $q$ are set as 100. In the following, we will evaluate the performance of our newly proposed method under a much more challenging setting: $p=q=500$. This is quite a large scale multiple testing case, because it is actually testing $500 \times 500 = 250,000$ hypotheses simultaneously. The results are summarized in Figures 7-12 and Table 2. Note that noodle method fails under this setting due to the computational complexity. Our sandwich method still performs well for approximating the true FDP, while PFA underestimates the true value. 

\begin{figure}[h!!!]
\begin{tabular}{cc}
  \includegraphics[width=50mm]{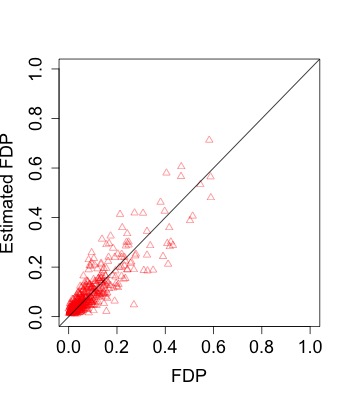} &   \includegraphics[width=50mm]{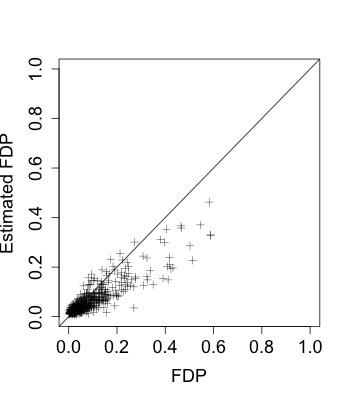} \\
\small{(a) $f(2,4), \mathbf{B} \sim  U(-1,1)$} & \small{(b) $f(2,4), \mathbf{B} \sim  U(-1,1) $} \\[6pt]
 \includegraphics[width=50mm]{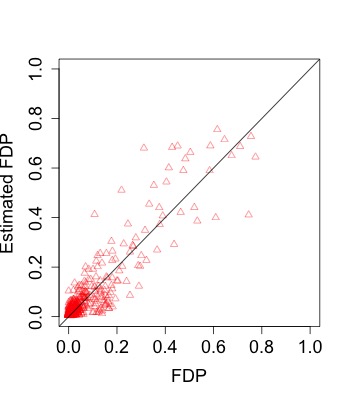} &   \includegraphics[width=50mm]{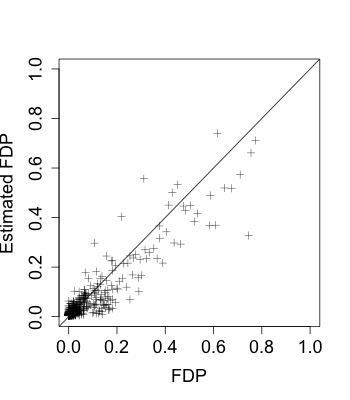} \\
\small{(c)$f(3,3), \mathbf{B} \sim  N(0,1)$} & \small{(d)$f(3,3), \mathbf{B} \sim  N(0,1)$} \\[6pt]
\end{tabular}
\centering
\caption{\footnotesize{Model 1, the estimated values of FDP obtained by noodle method (blue circle), sandwich method (red triangle) and PFA (black crossover). The first row shows the first setting of Model 1, and the second row shows the second setting of Model 1. Here, n=m=100, p=q=500, and t=0.0001.}}
\label{100500Model1}
\end{figure}

\begin{figure}[h!!!]
\begin{tabular}{cc}
  \includegraphics[width=50mm]{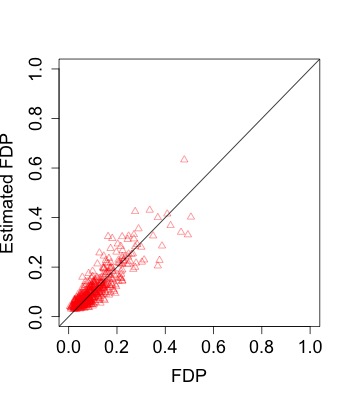} &   \includegraphics[width=50mm]{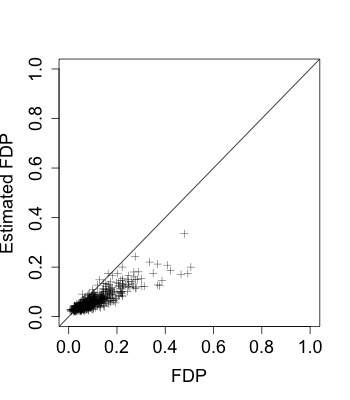} \\
\small{(a) $(\rho_1,\rho_2) = (0.5,0.3)$} & \small{(b) $(\rho_1,\rho_2) = (0.5,0.3)$} \\[6pt]
 \includegraphics[width=50mm]{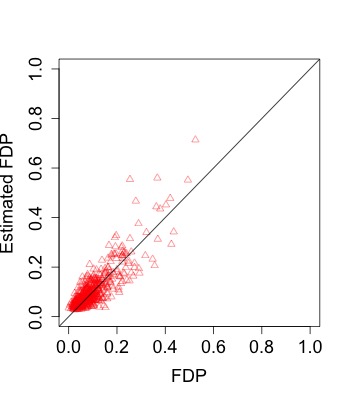} &   \includegraphics[width=50mm]{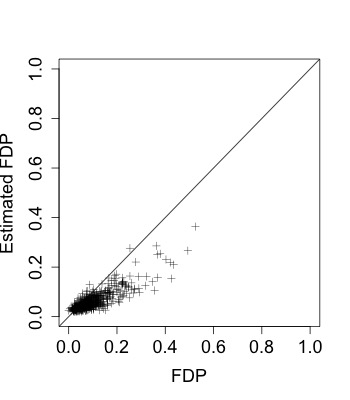} \\
\small{(c)$(\rho_1,\rho_2) = (0.5,0.8)$} & \small{(d)$(\rho_1,\rho_2) = (0.5,0.8)$} \\[6pt]
\end{tabular}
\centering
\caption{\footnotesize{Model 2, the estimated values of FDP obtained by noodle method (blue circle), sandwich method (red triangle) and PFA (black crossover). The first row shows the first setting of Model 2, and the second row shows the second setting of Model 2. Here, n=m=100, p=q=500, and t=0.0001.}}
\label{100500Model2}
\end{figure}

\begin{figure}[h!!!]
\begin{tabular}{cc}
  \includegraphics[width=50mm]{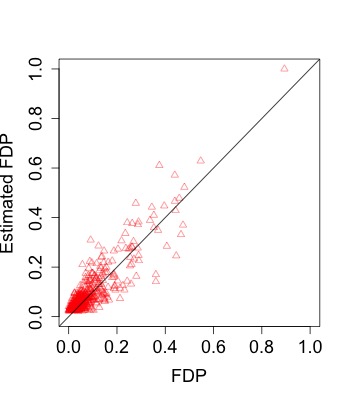} &   \includegraphics[width=50mm]{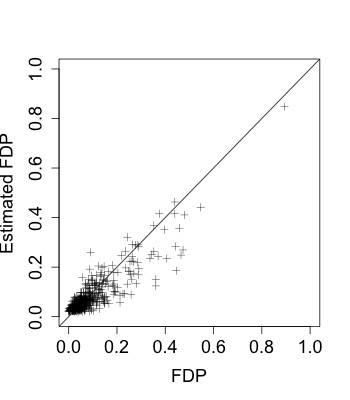} \\
\small{(a) $f(2,2), \mathbf{W} \sim Exp(1)$} 
& \small{(b) $f(2,2), \mathbf{W} \sim Exp(1)$} \\[6pt]
 \includegraphics[width=50mm]{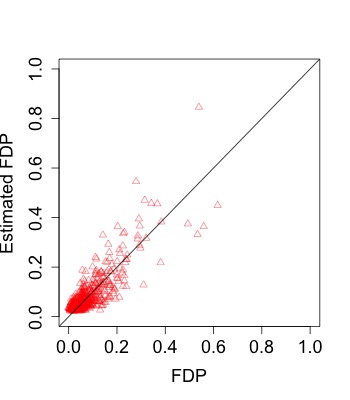} &   \includegraphics[width=50mm]{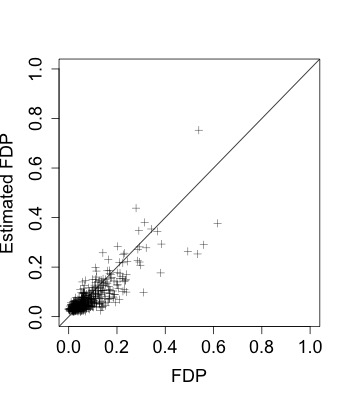} \\
\small{(c)$f(2,2), \mathbf{W} \sim \sqrt{\frac{2}{3}}t_6$} 
& \small{(d)$f(2,2), \mathbf{W} \sim \sqrt{\frac{2}{3}}t_6$} \\[6pt]
\end{tabular}
\centering
\caption{\footnotesize{Model 3: factor (2,2) case, the estimated values of FDP obtained by noodle method (blue circle), sandwich method (red triangle) and PFA (black crossover). The first row represents the case where $\mathbf{W}$ with each element following Exp(1) distribution, and the second row shows the case where $\mathbf{W}$ with each element following $\sqrt{\frac{2}{3}}t_6$ distribution. Here, n=m=100, p=q=500, and t=0.0001.}}
\label{100500Model3_f(2,2)}
\end{figure}

\begin{figure}[h!!!]
\begin{tabular}{cc}
  \includegraphics[width=50mm]{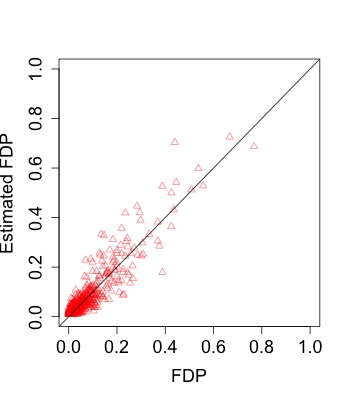} &   \includegraphics[width=50mm]{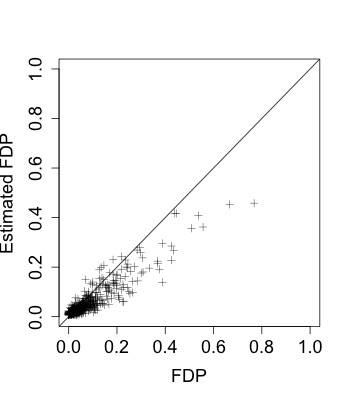} \\
\small{(a) $f(3,3), \mathbf{W} \sim Exp(1)$} 
& \small{(b) $f(3,3), \mathbf{W} \sim Exp(1)$} \\[6pt]
 \includegraphics[width=50mm]{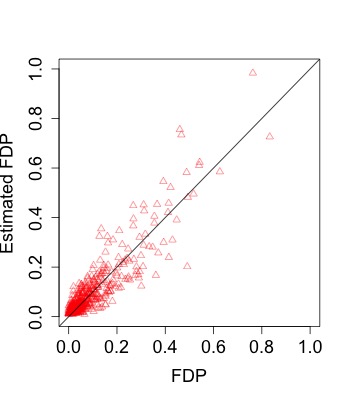} &   \includegraphics[width=50mm]{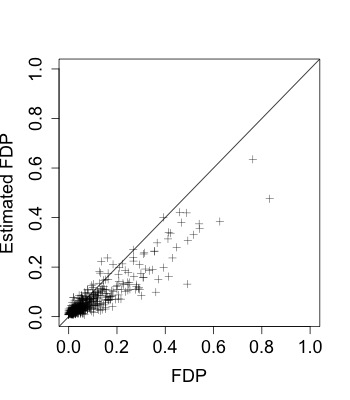} \\
\small{(c)$f(3,3), \mathbf{W} \sim \sqrt{\frac{2}{3}}t_6$} 
& \small{(d)$f(3,3), \mathbf{W} \sim \sqrt{\frac{2}{3}}t_6$} \\[6pt]
\end{tabular}
\centering
\caption{\footnotesize{Model 3: factor (3,3) case, the estimated values of FDP obtained by noodle method (blue circle), sandwich method (red triangle) and PFA (black crossover). The first row represents the case where $\mathbf{W}$ with each element following Exp(1) distribution, and the second row shows the case where $\mathbf{W}$ with each element following $\sqrt{\frac{2}{3}}t_6$ distribution. Here, n=m=100, p=q=500, and t=0.0001.}}
\label{100500Model3_f(3,3)}
\end{figure}

\begin{figure}[h!!!]
\begin{tabular}{cc}
  \includegraphics[width=50mm]{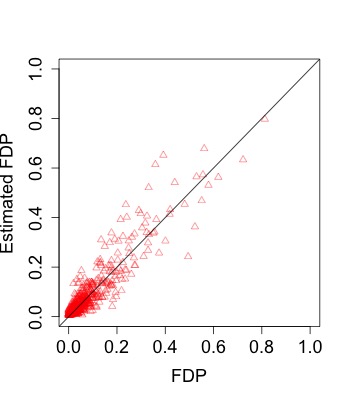} &   \includegraphics[width=50mm]{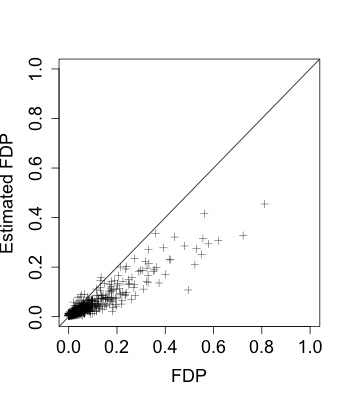} \\
\small{(a) $f(4,4), \mathbf{W} \sim Exp(1)$} 
& \small{(b) $f(4,4), \mathbf{W} \sim Exp(1)$} \\[6pt]
 \includegraphics[width=50mm]{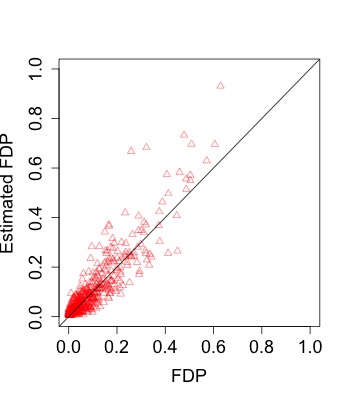} &   \includegraphics[width=50mm]{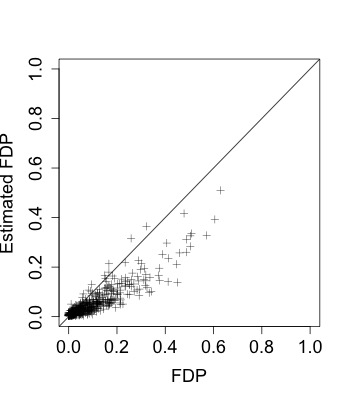} \\
\small{(c)$f(4,4), \mathbf{W} \sim \sqrt{\frac{2}{3}}t_6$} 
& \small{(d)$f(4,4), \mathbf{W} \sim \sqrt{\frac{2}{3}}t_6$} \\[6pt]
\end{tabular}
\centering
\caption{\footnotesize{Model 3: factor (4,4) case, the estimated values of FDP obtained by noodle method (blue circle), sandwich method (red triangle) and PFA (black crossover). The first row represents the case where $\mathbf{W}$ with each element following Exp(1) distribution, and the second row shows the case where $\mathbf{W}$ with each element following $\sqrt{\frac{2}{3}}t_6$ distribution. Here, n=m=100, p=q=500, and t=0.0001.}}
\label{100500Model3_f(4,4)}
\end{figure}

\begin{figure}[h!!!]
\begin{tabular}{cc}
  \includegraphics[width=50mm]{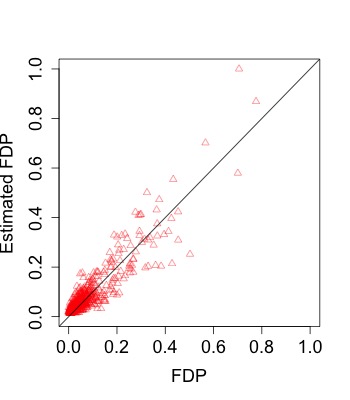} &   \includegraphics[width=50mm]{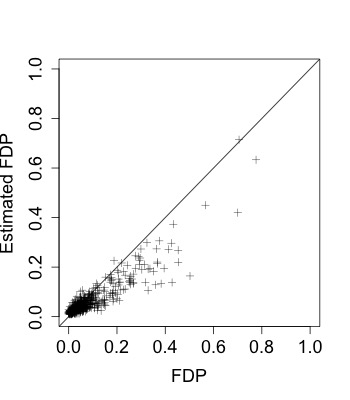} \\
\small{(a) $f(2,4), \mathbf{W} \sim Exp(1)$} 
& \small{(b) $f(2,4), \mathbf{W} \sim Exp(1)$} \\[6pt]
 \includegraphics[width=50mm]{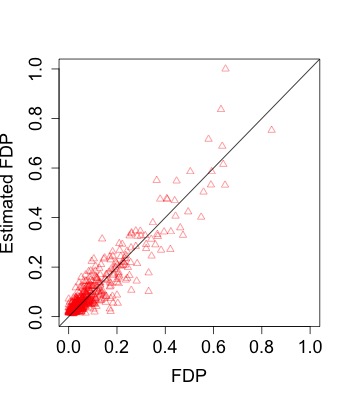} &   \includegraphics[width=50mm]{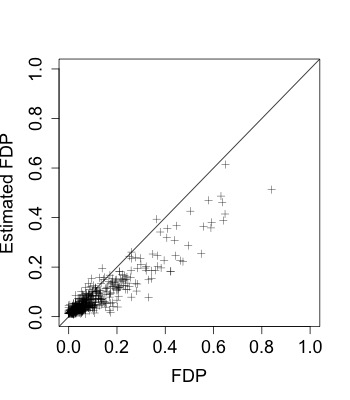} \\
\small{(c)$f(2,4), \mathbf{W} \sim \sqrt{\frac{2}{3}}t_6$} 
& \small{(d)$f(2,4), \mathbf{W} \sim \sqrt{\frac{2}{3}}t_6$} \\[6pt]
\end{tabular}
\centering
\caption{\footnotesize{Model 3: factor (2,4) case, the estimated values of FDP obtained by noodle method (blue circle), sandwich method (red triangle) and PFA (black crossover). The first row represents the case where $\mathbf{W}$ with each element following Exp(1) distribution, and the second row shows the case where $\mathbf{W}$ with each element following $\sqrt{\frac{2}{3}}t_6$ distribution. Here, n=m=100, p=q=500, and t=0.0001.}}
\label{100500Model3_f(2,4)}
\end{figure}

\begin{table}[!ht]
\centering
   {\caption{\footnotesize{\textit{Setting with n=m=50, p=q=100 and t=0.001} -- Mean and standard deviation of $\widehat{\FDP}(t)-\FDP(t)$. Results are presented in percent.}}\label{p=q=100}}
\begin{tabular}{|ccllcllcllc|}
\hline
\multicolumn{11}{|c|}{Results (\%) for the following methods:}                                                                                                                                                                                                                                         \\ \hline
                                      &                      & \multicolumn{1}{c}{}      & \multicolumn{1}{c}{}      &                      & \multicolumn{1}{c}{}      & \multicolumn{1}{c}{}      &                      & \multicolumn{1}{c}{}     & \multicolumn{1}{c}{}   &                       \\
n=m=50                                & \multirow{2}{*}{}    & \multicolumn{2}{c}{Method 1}                          &                      & \multicolumn{2}{c}{Method 2}                          &                      & \multicolumn{2}{c}{PFA}                           &                       \\ \cline{3-4} \cline{6-7} \cline{9-10}
p=q=100                               &                      & \multicolumn{1}{c}{Bias}  & \multicolumn{1}{c}{Sd}    &                      & \multicolumn{1}{c}{Bias}  & \multicolumn{1}{c}{Sd}    &                      & \multicolumn{1}{c}{Bias} & \multicolumn{1}{c}{Sd} &                       \\ \hline
                                      &                      & \multicolumn{1}{c}{}      & \multicolumn{1}{c}{}      &                      & \multicolumn{1}{c}{}      & \multicolumn{1}{c}{}      &                      & \multicolumn{1}{c}{}     & \multicolumn{1}{c}{}   &                       \\
\multicolumn{11}{|c|}{Model 1}                                                                                                                                                                                                                                                                         \\
\multicolumn{1}{|l}{}                 & \multicolumn{1}{l}{} &                           &                           & \multicolumn{1}{l}{} &                           &                           & \multicolumn{1}{l}{} &                          &                        & \multicolumn{1}{l|}{} \\
$f(2,4), B\sim U(-1,1)$               &                      & \multicolumn{1}{c}{0.885} & \multicolumn{1}{c}{2.583} &                      & \multicolumn{1}{c}{0.363} & \multicolumn{1}{c}{2.397} &                      & -1.931                   & 2.842                  &                       \\
$f(3,3), B\sim N(0,1)$                &                      & 1.141                     & 3.912                     &                      & 0.682                     & 3.541                     &                      & -0.807                   & 2.751                  &                       \\
\multicolumn{1}{|l}{}                 &                      & \multicolumn{1}{c}{}      & \multicolumn{1}{c}{}      &                      & \multicolumn{1}{c}{}      & \multicolumn{1}{c}{}      &                      & \multicolumn{1}{c}{}     & \multicolumn{1}{c}{}   &                       \\
\multicolumn{11}{|c|}{Model 2}                                                                                                                                                                                                                                                                         \\
\multicolumn{1}{|l}{}                 &                      & \multicolumn{1}{c}{}      & \multicolumn{1}{c}{}      &                      & \multicolumn{1}{c}{}      & \multicolumn{1}{c}{}      &                      & \multicolumn{1}{c}{}     & \multicolumn{1}{c}{}   &                       \\
$(\rho_1,\rho_2)=(0.5,0.3)$           &                      & 0.716                     & 2.360                     & \multicolumn{1}{l}{} & 0.206                     & 2.247                     & \multicolumn{1}{l}{} & -2.570                   & 2.594                  &                       \\
$(\rho_1,\rho_2)=(0.5,0.8)$           &                      & 0.818                     & 2.774                     &                      & 0.314                     & 2.671                     &                      & -1.318                   & 3.518                  &                       \\
\multicolumn{1}{|l}{}                 &                      & \multicolumn{1}{c}{}      & \multicolumn{1}{c}{}      &                      & \multicolumn{1}{c}{}      & \multicolumn{1}{c}{}      &                      & \multicolumn{1}{c}{}     & \multicolumn{1}{c}{}   &                       \\
\multicolumn{11}{|c|}{Model 3}                                                                                                                                                                                                                                                                         \\
\multicolumn{1}{|l}{}                 &                      & \multicolumn{1}{c}{}      & \multicolumn{1}{c}{}      &                      & \multicolumn{1}{c}{}      & \multicolumn{1}{c}{}      &                      & \multicolumn{1}{c}{}     & \multicolumn{1}{c}{}   &                       \\
$f(2,2), W\sim Exp(1)$                &                      & 1.090                     & 2.898                     & \multicolumn{1}{l}{} & 0.620                     & 2.564                     & \multicolumn{1}{l}{} & -0.686                   & 2.223                  &                       \\
$f(2,2), W\sim \sqrt{\frac{2}{3}}t_6$ &                      & 1.009                     & 3.169                     &                      & 0.480                     & 2.874                     &                      & -0.878                   & 2.564                  &                       \\
$f(2,4), W\sim Exp(1)$                &                      & 0.936                     & 2.568                     &                      & 0.392                     & 2.373                     &                      & -1.623                   & 2.822                  &                       \\
$f(2,4), W\sim \sqrt{\frac{2}{3}}t_6$ &                      & 0.931                     & 2.497                     &                      & 0.361                     & 2.347                     &                      & -1.847                   & 3.101                  &                       \\
$f(3,3), W\sim Exp(1)$                &                      & 0.977                     & 2.757                     &                      & 0.402                     & 2.541                     &                      & -1.841                   & 2.777                  &                       \\
$f(3,3), W\sim \sqrt{\frac{2}{3}}t_6$ &                      & 1.033                     & 2.697                     &                      & 0.436                     & 2.479                     &                      & -1.978                   & 3.117                  &                       \\
$f(4,4), W\sim Exp(1)$                &                      & 0.950                     & 2.867                     &                      & 0.366                     & 2.688                     &                      & -2.682                   & 3.514                  &                       \\
$f(4,4), W\sim \sqrt{\frac{2}{3}}t_6$ &                      & 1.176                     & 2.634                     &                      & 0.569                     & 2.426                     &                      & -2.576                   & 3.402                  &                       \\
\multicolumn{11}{|l|}{}                                                                                                                                                                                                                                                                                \\ \hline
\end{tabular}
\end{table}

\begin{table}[h!!!]
\centering
   {\caption{\footnotesize{\textit{Setting with n=m=100, p=q=500 and t=0.0001} -- Mean and standard deviation of $\widehat{FDP}(t)-FDP(t)$. Results are presented in percent.}}\label{p=q=500}}
\begin{tabular}{|cclllllc|}
\hline
\multicolumn{8}{|c|}{Results (\%) for the following methods:}                                                                                                                                                       \\ \hline
                                      &                      & \multicolumn{1}{c}{}     & \multicolumn{1}{c}{}   & \multicolumn{1}{c}{} & \multicolumn{1}{c}{}     & \multicolumn{1}{c}{}   &                       \\
n=m=50                                & \multirow{2}{*}{}    & \multicolumn{2}{c}{Method 2}                      & \multicolumn{1}{c}{} & \multicolumn{2}{c}{PFA}                           &                       \\ \cline{3-4} \cline{6-7}
p=q=100                               &                      & \multicolumn{1}{c}{Bias} & \multicolumn{1}{c}{Sd} & \multicolumn{1}{c}{} & \multicolumn{1}{c}{Bias} & \multicolumn{1}{c}{Sd} &                       \\ \hline
                                      &                      & \multicolumn{1}{c}{}     & \multicolumn{1}{c}{}   & \multicolumn{1}{c}{} & \multicolumn{1}{c}{}     & \multicolumn{1}{c}{}   &                       \\
\multicolumn{8}{|c|}{Model 1}                                                                                                                                                                                       \\
\multicolumn{1}{|l}{}                 & \multicolumn{1}{l}{} &                          &                        &                      &                          &                        & \multicolumn{1}{l|}{} \\
$f(2,4), B\sim U(-1,1)$               &                      & 0.454                    & 4.813                  &                      & -2.459                   & 5.239                  &                       \\
$f(3,3), B\sim N(0,1)$                &                      & 0.645                    & 5.715                  &                      & -1.009                   & 5.124                  &                       \\
\multicolumn{1}{|l}{}                 &                      & \multicolumn{1}{c}{}     & \multicolumn{1}{c}{}   & \multicolumn{1}{c}{} & \multicolumn{1}{c}{}     & \multicolumn{1}{c}{}   &                       \\
\multicolumn{8}{|c|}{Model 2}                                                                                                                                                                                       \\
\multicolumn{1}{|l}{}                 &                      & \multicolumn{1}{c}{}     & \multicolumn{1}{c}{}   & \multicolumn{1}{c}{} & \multicolumn{1}{c}{}     & \multicolumn{1}{c}{}   &                       \\
$(\rho_1,\rho_2)=(0.5,0.3)$           &                      & 0.192                    & 3.882                  &                      & -4.201                   & 4.867                  &                       \\
$(\rho_1,\rho_2)=(0.5,0.8)$           &                      & 0.648                    & 4.507                  &                      & -3.839                   & 4.759                  &                       \\
\multicolumn{1}{|l}{}                 &                      & \multicolumn{1}{c}{}     & \multicolumn{1}{c}{}   & \multicolumn{1}{c}{} & \multicolumn{1}{c}{}     & \multicolumn{1}{c}{}   &                       \\
\multicolumn{8}{|c|}{Model 3}                                                                                                                                                                                       \\
\multicolumn{1}{|l}{}                 &                      & \multicolumn{1}{c}{}     & \multicolumn{1}{c}{}   & \multicolumn{1}{c}{} & \multicolumn{1}{c}{}     & \multicolumn{1}{c}{}   &                       \\
$f(2,2), W\sim Exp(1)$                &                      & 1.008                    & 4.712                  &                      & -0.771                   & 4.577                  &                       \\
$f(2,2), W\sim \sqrt{\frac{2}{3}}t_6$ &                      & 0.421                    & 4.810                  &                      & -1.082                   & 4.615                  &                       \\
$f(2,4), W\sim Exp(1)$                &                      & 0.690                    & 4.607                  &                      & -2.201                   & 5.021                  &                       \\
$f(2,4), W\sim \sqrt{\frac{2}{3}}t_6$ &                      & 0.358                    & 4.966                  &                      & -3.072                   & 5.742                  &                       \\
$f(3,3), W\sim Exp(1)$                &                      & 0.595                    & 4.284                  &                      & -2.153                   & 4.436                  &                       \\
$f(3,3), W\sim \sqrt{\frac{2}{3}}t_6$ &                      & 0.538                    & 5.312                  &                      & -2.720                   & 5.541                  &                       \\
$f(4,4), W\sim Exp(1)$                &                      & 0.602                    & 4.736                  &                      & -3.987                   & 5.953                  &                       \\
$f(4,4), W\sim \sqrt{\frac{2}{3}}t_6$ &                      & 0.806                    & 5.571                  &                      & -4.275                   & 5.640                  &                       \\
\multicolumn{8}{|l|}{}                                                                                                                                                                                              \\ \hline
\end{tabular}
\end{table} 

\section{Data Analysis}

Electroencephalogram (EEG) has been widely considered as an effective approach for detecting spontaneous fluctuations in brain activity. We will illustrate our newly proposed multiple testing procedures (both noodle method and sandwich method) on an EEG data set from a study to examine EEG correlating of genetic predisposition to alcoholism. We will compare two groups of subjects, alcoholic and control, to study the influence of alcohol on the brain. For each subject, the data contains measurements from 64 electrodes placed on the scalp sampled at 256 Hz for 1 second, while the subjects were performing a visual object recognition task. The data set and the more detailed description can be accessed via \textit{kdd.ics.uci.edu/databases/eeg}. 

In our study, let $\mathbf{Y}_1,...,\mathbf{Y}_n, n=77$, denote the voltage (in micro volts) for the group of alcoholic subjects, and $\mathbf{Z}_1,...,\mathbf{Z}_m, m=45$, denote the voltage for the group of control subjects. Thus, each sample contains $p \times q = 64 \times 256 = 16384$ values of measurements. We further assume that the voltage of the two groups on each subject are from two matrix normal distributions with possibly different mean matrix but the same column covariance matrix $\mathbf{U}$ and row covariance matrix $\mathbf{V}$. More specifically, $\mathbf{Y}_i \sim \mathcal{MN}(\mathbf{\mu}^y,\mathbf{U},\mathbf{V})$ for $i=1,...,77$ and $\mathbf{Z}_j \sim \mathcal{MN}(\mathbf{\mu}^z,\mathbf{U},\mathbf{V})$ for $j=1,...,45$. The matrix normal assumption has been supported by the previous studies on this EEG data for some other statistical problems, see Li, Kim and Altman (2010) and Xia and Li (2017). In our problem, testing whether EEG correlating of genetic predisposition to alcoholism can be formulated as a multiple hypothesis testing problem on $H_{0ij}:\mathbf{\mu}^y_{ij}=\mathbf{\mu}^z_{ij}$ \textit{versus} $H_{1ij}:\mathbf{\mu}^y_{ij}\neq \mathbf{\mu}^z_{ij}$ for $i=1,...,64, j=1,...,256$.

\begin{figure}[h!!!]
\begin{tabular}{cc}
  \includegraphics[width=80mm]{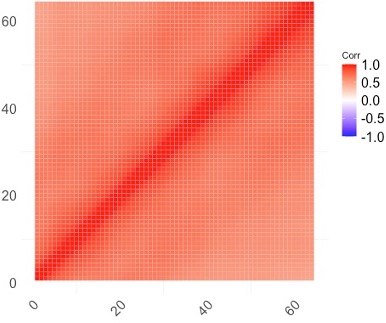} &   \includegraphics[width=80mm]{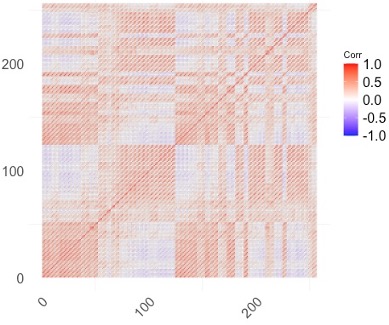} \\
\small{(a) $\hat{\Sigma}_{64 \times 64}$} & \small{(b) $\hat{\Sigma}_{256 \times 256}$ } \\[6pt]
\end{tabular}
\centering
\caption{\footnotesize{Heat map of the two estimated correlation matrix from EEG data.}}
\label{heatmap}
\end{figure}

\begin{figure}[h!!!]
\begin{tabular}{ccc}
  \includegraphics[width=50mm]{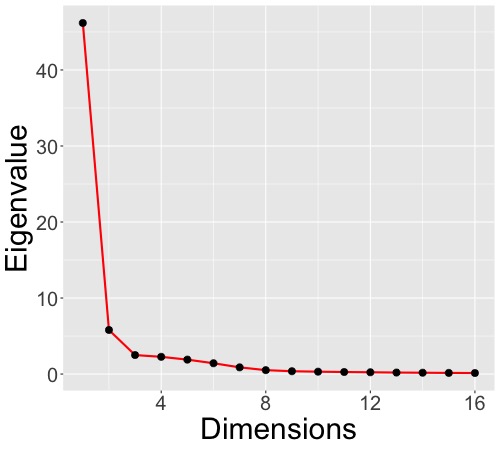} &   \includegraphics[width=50mm]{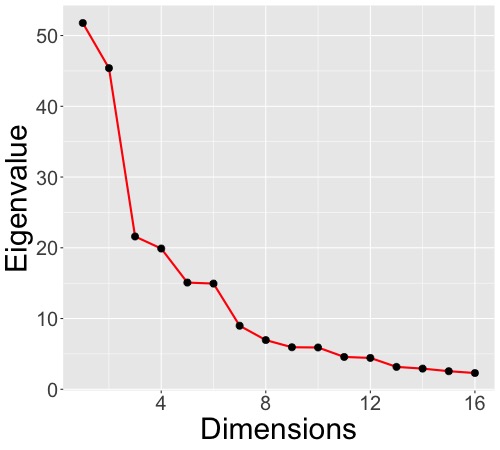} &
  \includegraphics[width=50mm]{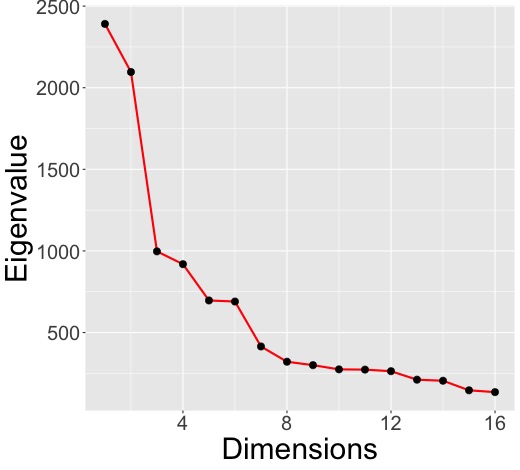}\\
\small{(a) $\mathbb{\lambda}$ from $corr(\hat{\Sigma}_{64 \times 64})$} & 
\small{(b) $\mathbf{\xi}$ from $corr(\hat{\Sigma}_{256 \times 256}$) } &
\small{(c) $\mathbb{\lambda} \otimes\mathbb{\xi}$ }\\[6pt]
\end{tabular}
\centering
\caption{\footnotesize{Plot of eigenvalues for two estimated correlation matrix and their sorted product from EEG data, $\lambda=(\lambda_1,...,\lambda_p)$, $\xi=(\xi_1,...,\xi_q)$.}}
\label{eigenvalue}
\end{figure}

The EEG data reflects the brain electric activity in an spatial-temporal pattern, where the dependence from either direction should not be simply ignored. The temporal correlation is easier to understand, as the activities of the same brain regions are recorded through time. The spatial correlation has a deeper scientific foundation, reflecting the brain functional connectivity (Fox and Raicle 2007). Figure \ref{heatmap} shows the heat map of the two estimated correlation matrices. Since the dimensionality $p$ and $q$ are not high compared with the sample size, we would expect that the two estimated correlation matrices are not deviating away from the population correlation matrices much. Nevertheless, Figure \ref{heatmap} does not show a clear pattern in the two matrices. Figure \ref{eigenvalue} further shows the scree plots of the two estimated correlation matrices where the first few eigenvalues are substantially larger than the others,  demonstrating the strong dependence in both matrices. This falls within the framework of our propose methods.  

\begin{figure}[h!!!]
\begin{tabular}{cc}
  \includegraphics[width=80mm]{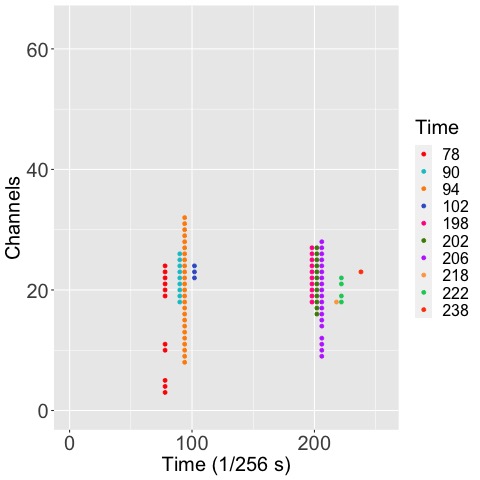} &   \includegraphics[width=80mm]{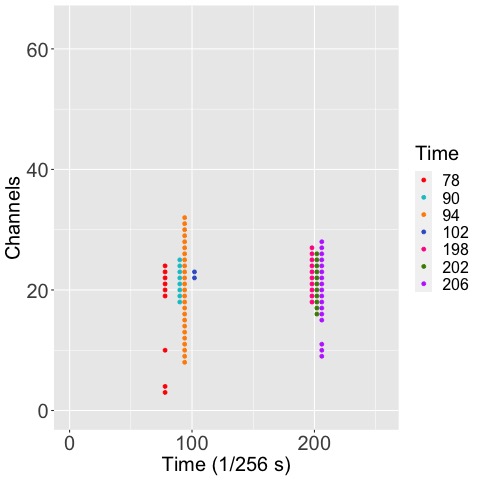} \\
\small{(a) $q=0.2$} & \small{(b) $q=0.1$ } \\[6pt]
\includegraphics[width=80mm]{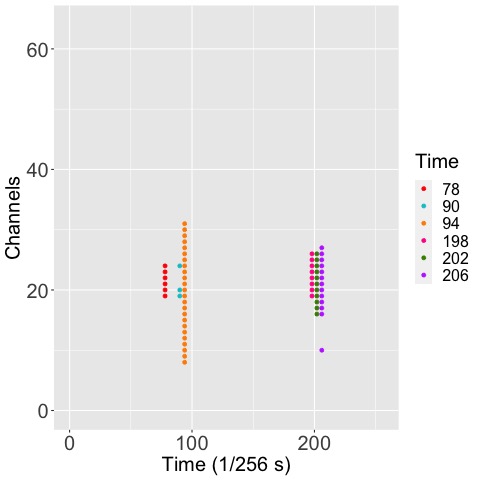} &   \includegraphics[width=80mm]{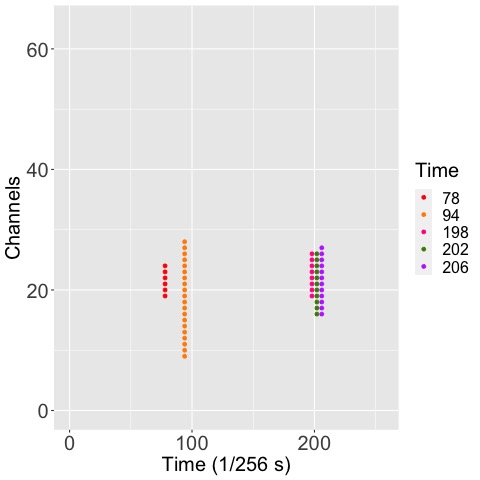} \\
\small{(c)} $q=0.01$& \small{(d) $q=0.001$} \\[6pt]
\end{tabular}
\centering
\caption{\footnotesize{Plots of the selected hypotheses with different threshold values q: (a) 0.2 (b) 0.1 (c) 0.01 (d) 0.001}}
\label{plot1}
\end{figure}

Figure \ref{plot1} shows the results of selected signals, by rejecting the hypotheses where the corresponding p-values are no greater than a predetermined threshold value $q$. Both noodle method and sandwich method return the same results here. By choosing a larger threshold value, more signals will be detected, whereas a smaller threshold value will lead to fewer discoveries. Although channels discovered are different according to different threshold values, the time when a signal is discovered is relatively stable. It is also interesting that the time lag between signals may reflect the causal effect or the direction of influence of the regional brain activities with response to the task. 

\begin{figure}[h!!!]
\begin{tabular}{cc}
  \includegraphics[width=70mm]{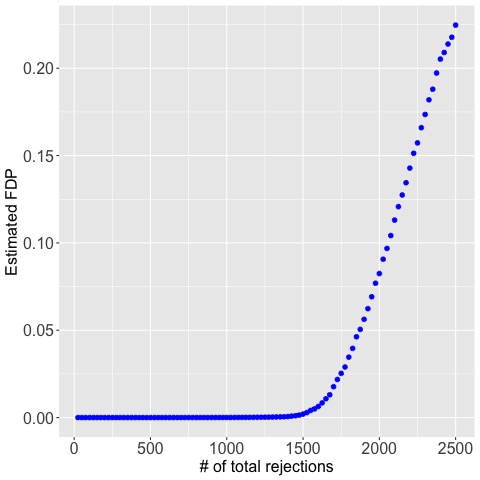} &   \includegraphics[width=70mm]{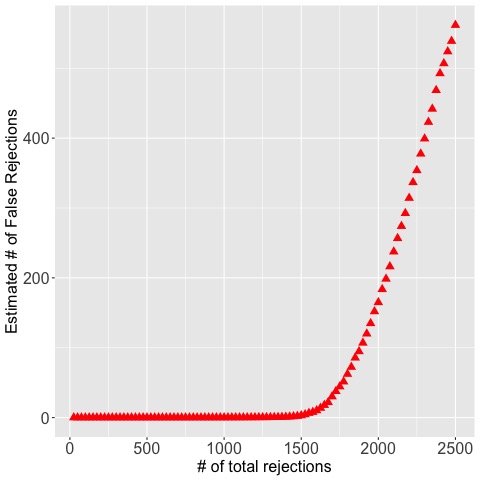} \\
\small{(a)} & \small{(b) } \\[6pt]
 \end{tabular}
\centering
\caption{\footnotesize{(a) Estimated $\FDP$ and (b) estimated number of false discoveries for p=64 channels and q=256 Hz}}
\label{plot2}
\end{figure}

The p-value strategy encourages us to select signals with smaller p-values. For different threshold values, the selection can be quite different, so is the FDP approximation. Practitioners are also interested in the false discovery inference according to the total number of rejections. We sort the p-values in increasing order, denoted as $(P_{(1)},...,P_{(16384)})$, and then consider a sequence of threshold values $t_i=P_{(i \times 25)}$, $i=1,...,100$. The total rejections are computed as $R_i(t)=\sum_{j=1}^{p\times q} I(P_j \leq t_i)$. In Figure \ref{plot2}, we report the estimated FDP and the number of false discoveries versus the number of total rejections. The monotonicity pattern in both plots is not surprising, as more rejections will involve more false rejections. However, when the total number of rejections is less than 1500, both the estimated FDP and the number of false discoveries are very close to 0, suggesting that the hypotheses rejected during this stage are very likely to yield the true signals.

Our methods have shed some new light on the significance detection for the large scale two sample comparison. In the current paper, the methods are based on a sample covariance matrix estimator. This approach is general, as we do not require a particular structure of the two population covariance matrices. When a priori knowledge for the brain function connectivity is available, we can choose a better estimator for the covariance matrices, and the corresponding FDP approximation can be further improved. Another limitation is the matrix normal assumption, which imposes a structured dependence from the row and column correlations. Such assumption may be violated in practice, even if each element in the matrix data follows a normal distribution. One possibility is to detect significance by rows, and then by columns. The final selection can be the intersection from row selection and column selection. We would like to explore this issue in our future research. By and large, the current paper provides an effective step for the significance detection in the large scale matrix valued data, which will be useful for the brain related research.

\section{Appendix}

\begin{lemma}  For any matrix $\hat{\bSigma}$ with eigenvalues $\{\lambda_i\}$ (in non-increasing order) and the corresponding eigenvectors $\{\bgamma_i\}$. Let $\widehat{\bSigma}$ be an estimate of $\bSigma$, with corresponding eigenvalues $\{\widehat{\lambda}_i\}$ and eigenvectors $\{\widehat{\bgamma}_i\}$. We have
\begin{equation*}
|\widehat{\lambda}_i-\lambda_i|\leq\|\widehat{\bSigma}-\bSigma\| \quad
\mbox{and} \quad \|\widehat{\bgamma}_i-\bgamma_i\|\leq\frac{\sqrt{2}\|\widehat{\bSigma}-\bSigma\|}
{\min(|\widehat{\lambda}_{i-1}-\lambda_i|,|\lambda_i-\widehat{\lambda}_{i+1}|)}.
\end{equation*}
\end{lemma}
The first result is referred to Weyl's Theorem (Horn \& Johnson, 1990) and the second result is called the $\sin\theta$ Theorem (Davis \& Kahan, 1970). 

\noindent{\bf Proof of Proposition 1:}

We further denote those diagonal elements of $\bT^{1/2}$ as $\{\sqrt{T_i}\}_{i=1}^{pq}$, then 
\begin{eqnarray*}
vec(\bX)|\{T_l\}_{l=1}^{pq}&\sim& N_{pq}(\bT^{1/2}vec(\bmu^{\star}), \bT^{1/2}\bSigma_2\otimes\bSigma_1\bT^{1/2})\\
T_i&\sim&InverseGamma(\frac{n+m-2}{2},\frac{n+m-2}{2}), \quad\quad i=1,\cdots, pq
\end{eqnarray*}
where we let $\bmu^{\star}=\sqrt{\frac{nm}{n+m}}(\bmu_y-\bmu_z)\circ\bSigma$. 

Let 
\begin{equation*}
\Delta\equiv\frac{1}{pq}\sum_{l\in\text{true null}}\bI(P_i\leq t|\bW)-\frac{1}{pq}\sum_{l\in\{\text{true null}\}}[\Phi(a_l(z_{t/2}+\zeta_l))+\Phi(a_l(z_{t/2}-\zeta_l))]. 
\end{equation*}
We want to show that $\Delta=O_p((pq)^{-\delta/2})+O_p((n+m-2)^{-1/2})$. To prove this, it suffices to show that 
\begin{equation}\label{delta1}
|\frac{1}{pq}\sum_{i\in\{\text{true null}\}}\bI(P_l\leq t|\bW)-\frac{1}{pq}\sum_{l\in\{\text{true null}\}}P(P_i\leq t|\bW)|=O_p((pq)^{-\delta/2})+O_p((n+m-2)^{-1/2})
\end{equation}
and 
\begin{equation}\label{delta2}
|\frac{1}{pq}\sum_{i\in\{\text{true null}\}}P(P_i\leq t|\bW)-\frac{1}{pq}\sum_{i\in\{\text{true null}\}}[\Phi(a_l(z_{t/2}+\zeta_l))+\Phi(a_l(z_{t/2}-\zeta_l))]|=O((n+m-2)^{-1/2}). 
\end{equation}

To show (\ref{delta1}), note that 
\begin{eqnarray*}
&&Var((pq)^{-1}\sum_{l\in\{\text{true null}\}}\bI(P_l\leq t|\bW))\\
&=&\frac{1}{(pq)^2}\sum_{i\in\{\text{true null}\}}Var(\bI(P_i\leq t|\bW))+\frac{1}{(pq)^2}\sum_{i,j\in\{\text{true null}\}, i\neq j}cov(\bI(P_i\leq t|\bW),\bI(P_j\leq t|\bW))\\
&=&O((pq)^{-1})+O((pq)^{-1}(\sum_{l=h+1}^{pq}\theta_l^2)^{1/2})+O((n+m-2)^{-1})
\end{eqnarray*}
based on the proof of Theorem 2 in Fan and Han (2017). By the condition in Proposition 1, $(pq)^{-1}(\sum_{l=h+1}^{pq}\theta_l^2)^{1/2}=O((pq)^{-\delta})$. Therefore, expression (\ref{delta1}) is $O_p((pq)^{-\delta/2})+O_p((n+m-2)^{-1/2})$. By Lemma 4 in Fan and Han (2017), the conclusion in expression (\ref{delta2}) is also correct. Combining with the condition that $R(t)^{-1}=O_p((pq)^{-(1-\theta_1)})$, the proof is now complete. 

\noindent{\bf Proof of Theorem 1: }

Define an infeasible estimator $\widetilde{\bW}_2=(\bF^T\bF)^{-1}\bF^Tvec(\bX)$. Denote $\FDP_2(t)$ as the formula in $\FDP_{A,1}(t)$ with using the infeasible estimator $\widetilde{\bW}_2$. Define the infeasible estimator $\widetilde{\bW}_1=(\bF^T\bF)^{-1}\bF^Tvec(\widetilde{\bX})$. Denote $\FDP_1(t)$ as the formula in $\FDP_{A,1}(t)$ by using the infeasible estimator $\widetilde{\bW}_1$. Then we have 
\begin{equation*}
\widehat{\FDP}_1(t)-\FDP_{A,1}(t)=\widehat{\FDP}_1(t)-\FDP_2(t)+\FDP_2(t)-\FDP_1(t)+\FDP_1(t)-\FDP_{A,1}(t). 
\end{equation*}
We will first evaluate $\widehat{\FDP}_1(t)-\FDP_2(t)$. Define
\begin{eqnarray*}
\Delta_1&=&\sum_{l=1}^{pq}[\Phi(\widehat{a}_l(z_{t/2}+\widehat{\bff}_l^T\widehat{\bW}))-\Phi(a_l(z_{t/2}+\bff_l^T\widetilde{\bW}_2))], \\
\Delta_2&=&\sum_{l=1}^{pq}[\Phi(\widehat{a}_l(z_{t/2}-\widehat{\bff}_l^T\widehat{\bW}))-\Phi(a_l(z_{t/2}-\bff_l^T\widetilde{\bW}_2))]. 
\end{eqnarray*}
Then we have $\widehat{\FDP}_1(t)-\FDP_2(t)=R(t)^{-1}[\Delta_1+\Delta_2]$. 

We further let $\Delta_1=\sum_{l=1}^{pq}\Delta_{1l}$, where 
\begin{eqnarray*}
\Delta_{1l}&=&\Phi(\widehat{a}_l(z_{t/2}+\widehat{\bff}_l^T\widehat{\bW}))-\Phi(\widehat{a}_l(z_{t/2}+\bff_l^T\widetilde{\bW}_2))\\
      &&+\Phi(\widehat{a}_l(z_{t/2}+\bff_l^T\widetilde{\bW}_2))-\Phi(a_l(z_{t/2}+\bff_l^T\widetilde{\bW}_2))\\
      &\equiv&\Delta_{11l}+\Delta_{12l}, 
\end{eqnarray*}
where the first part focuses on the difference between $\widehat{\bff}_l^T\widehat{\bW}$ and $\bff_l^T\widetilde{\bW}_2$ and the second part focuses on the difference between $\widehat{a}_l$ and $a_l$. 

For $\Delta_{11l}$, by the mean value theorem, there exists $\zeta_l$ between $\widehat{\bff}_l^T\widehat{\bW}$ and $\bff_l^T\widetilde{\bW}_2$ such that 
$\Delta_{11l}=\phi(\widehat{a}_l(z_{t/2}+\zeta_l))\widehat{a}_l(\widehat{\bff}_l^T\widehat{\bW}-\bff_l^T\widetilde{\bW}_2)$. By the condition that $\widehat{a}_l$ is bounded and so is $\phi(\widehat{a}_l(z_{t/2}+\zeta_l))\widehat{a}_l$. Then 
\begin{eqnarray*}
\sum_{l=1}^{pq}|\widehat{\bff}_l^T\widehat{\bW}-\bff_l^T\widetilde{\bW}_2|&=&\bone^T|\widehat{\bF}\widehat{\bW}-\bF\widetilde{\bW}_2|\\
   &=&\bone^T|\sum_{k=1}^h(\widehat{\brho}_k\widehat{\brho}_k^T-\brho_k\brho_k^T)vec(\bX)|\\
   &\leq&\sqrt{pq}\|\sum_{k=1}^h(\widehat{\brho}_k\widehat{\brho}_k^T-\brho_k\brho_k^T)\|\cdot\|vec(\bX)\|. 
\end{eqnarray*}
For the eigenvectors, 
\begin{equation*}
\|\sum_{k=1}^h(\widehat{\brho}_k\widehat{\brho}_k^T-\brho_k\brho_k^T)\|\leq\sum_{k=1}^h[\|\widehat{\brho}_k(\widehat{\brho}_k-\brho_k)^T\|+\|(\widehat{\brho}_k-\brho_k)\widehat{\brho}_k^T\|]\leq2\sum_{k=1}^h\|\widehat{\brho}_k-\brho_k\|
\end{equation*}
For $\|vec(\bX)\|$, note that $vec(\bX)=\bT^{1/2}vec(\widetilde{\bX})$, where $\bT^{1/2}=\diag(\sqrt{T_i})$ and 
\begin{equation*}
T_i\sim InverseGamma(\frac{n+m-2}{2},\frac{n+m-2}{2}) 
\end{equation*}
for $i=1,\cdots,pq$ independent of $vec(\widetilde{\bX})$. Therefore, 
\begin{equation*}
E\|vec(\bX)\|^2=\sum_{i=1}^{pq}E(T_i(vec(\bX))_i^2)=\sum_{i=1}^{pq}ET_iE[vec(\bX)]_i^2. 
\end{equation*}
Since $ET_i=\frac{n+m-2}{n+m-4}$ and $E[vec(\bX)]_i^2=(vec(\bmu^{\star}))_i^2+1$. Therefore, 
\begin{equation*}
E\|vec(\bX)\|^2=\frac{n+m-2}{n+m-4}(\|vec(\bmu^{\star})\|^2+pq). 
\end{equation*}
This implies that $\|vec(\bX)\|=O_p(\|vec(\bmu^{\star})\|+(pq)^{1/2})$. 

Next we evaluate $\Delta_{12l}$. By the mean value theorem, there exists $a_l^*\in(a_l, \widehat{a}_l)$ such that $\Delta_{12l}=\phi(a_l^*(z_{t/2}+\bff_l^T\widetilde{\bW}_2))(\widehat{a}_l-a_l)(z_{t/2}+\bff_l^T\widetilde{\bW}_2)$. Since both $a_l$ and $\widehat{a}_l$ are greater than 1, we have $a_l^*>1$ and hence $\phi(a_l^*(z_{t/2}+\bff_l^T\widetilde{\bW}))|z_{t/2}+\bff_l^T\widetilde{\bW}|$ is bounded. Therefore,
\begin{equation*}
|\sum_{l=1}^{pq}\Delta_{12l}|\leq C_1\sum_{l=1}^{pq}|\widehat{a}_l-a_l|
\end{equation*}
for some positive constant $C_1$. 

For the difference between $\widehat{a}_l$ and $a_l$, apply the mean value theorem, we have $|\widehat{a}_l-a_l|\leq C_2(\|\widehat{\bff}_l\|^2-\|\bff_l\|^2)$. We define $\brho_k=(\rho_{1k},\cdots,\rho_{pq,k})^T$ and $\widehat{\brho}_k=(\widehat{\rho}_{1k},\cdots,\widehat{\rho}_{pq,k})^T$, then 
\begin{eqnarray*}
\sum_{l=1}^{pq}[\|\widehat{\bff}_l\|^2-\|\bff_l\|^2]&=&\sum_{l=1}^{pq}|\sum_{k=1}^h(\widehat{\theta}_k-\theta_k)\widehat{\rho}_{lk}^2+\sum_{k=1}^h\theta_k(\widehat{\rho}_{lk}^2-\rho_{lk}^2)|\\
&\leq&\sum_{k=1}^h|\widehat{\theta}_k-\theta_k|\sum_{l=1}^{pq}\widehat{\rho}_{lk}^2+\sum_{k=1}^h\theta_k\sum_{l=1}^{pq}|\widehat{\rho}_{lk}^2-\rho_{lk}^2|. 
\end{eqnarray*}
Note that $\sum_{l=1}^{pq}\widehat{\rho}_{lk}^2=1$ by the definition of eigenvectors. 

Furthermore, we have
\begin{eqnarray*}
\sum_{l=1}^{pq}|\widehat{\rho}_{lk}^2-\rho_{lk}^2|&\leq&\{\sum_{l=1}^{pq}(\widehat{\rho}_{lk}-\rho_{lk})^2\sum_{l=1}^{pq}(\widehat{\rho}_{lk}+\rho_{lk})^2\}^{1/2}\\
  &\leq&\|\widehat{\brho}_k-\brho_k\|\{2\sum_{l=1}^{pq}(\widehat{\rho}_{lk}^2+\rho_{lk}^2)\}^{1/2}\\
  &=&2\|\widehat{\brho}_k-\brho_k\|. 
\end{eqnarray*}
Therefore, 
\begin{equation*}
|\sum_{l=1}^{pq}\Delta_{12l}|\leq C_3(\sum_{k=1}^h\theta_k\|\widehat{\brho}_k-\brho_k\|+|\widehat{\theta}_k-\theta_k|). 
\end{equation*}
To summarize, 
\begin{equation}\label{delta1}
|\Delta_1|\leq C_4(pq)^{1/2}\sum_{k=1}^h\|\widehat{\brho}_k-\brho_k\|O_p(\|vec(\bmu^{\star})\|+(pq)^{1/2})+C_5\sum_{k=1}^h[\theta_k\|\widehat{\brho}_k-\brho_k\|+|\widehat{\theta}_k-\theta_k|]. 
\end{equation}

Next we evaluate $|\FDP_2(t)-\FDP_1(t)|$. Note that $|\FDP_2(t)-\FDP_1(t)|=|\sum_{l=1}^{pq}\Delta_{3l}+\Delta_{4l}|$ where 
\begin{eqnarray*}
\Delta_{3l}&=&\Phi(a_l(z_{t/2}+\bff_l^T\widetilde{\bW}_2))-\Phi(a_l(z_{t/2}+\bff_l^T\widetilde{\bW}_1))\\
\Delta_{4l}&=&\Phi(a_l(z_{t/2}-\bff_l^T\widetilde{\bW}_2))-\Phi(a_l(z_{t/2}-\bff_l^T\widetilde{\bW}_1)). 
\end{eqnarray*}
To analyze $\Delta_{3l}$, apply the mean value theorem, there exists $\psi_l$ between $\bff_l^T\widetilde{\bW}_2$ and $\bff_l^T\widetilde{\bW}_1$ such that $\Delta_{3l}=\phi(a_l(z_{t/2}+\psi_l))a_l\bff_l^T(\widetilde{\bW}_2-\widetilde{\bW}_1)$. By the condition that $a_l$ is bounded and so is $\phi(a_l(z_{t/2}+\psi_l))a_l$. 
\begin{eqnarray*}
\sum_{l=1}^{pq}|\bff_l^T(\widetilde{\bW}_2-\widetilde{\bW}_1)|&=&\bone^T|\bF(\widetilde{\bW}_2-\widetilde{\bW}_1)|\\
&=&\bone^T|\sum_{k=1}^h\brho_k\brho_k^T(vec(\bX)-vec(\widetilde{\bX}))|\\
&\leq&\sqrt{pq}\|\sum_{k=1}^h\brho_k\brho_k^T\|\cdot\|vec(\bX)-vec(\widetilde{\bX})\|. 
\end{eqnarray*}
For the second term in the last line, $\|\sum_{k=1}^h\brho_k\brho_k^T\|=1$. For the third term in the last line, note that $vec(\bX)=\bV^{1/2}vec(\widetilde{\bX})$. Therefore, 
\begin{equation*}
E\|vec(\bX)-vec(\widetilde{\bX})\|^2=\sum_{l=1}^{pq}E(\sqrt{T_i}-1)^2(vec(\widetilde{\bX}))_i^2=\sum_{l=1}^{pq}E(\sqrt{T_i}-1)^2E(vec(\widetilde{\bX}))_i^2. 
\end{equation*}
Since $E(\sqrt{T_i}-1)^2=O((n+m-2)^{-1})$, we have
\begin{equation*}
\|vec(\bX)-vec(\widetilde{\bX})\|=O_p((n+m-2)^{-1/2}(\|vec(\bmu^{\star})\|+(pq)^{1/2})). 
\end{equation*}

We can apply similar analysis to $\Delta_{4l}$. Therefore, we have 
\begin{equation*}
|\FDP_2(t)-\FDP_1(t)|=\frac{1}{R(t)}O_p((n+m-2)^{-1/2}(pq)^{1/2}(\|vec(\bmu^{\star})\|+(pq)^{1/2})). 
\end{equation*}

Next we want to analyze $|\FDP_1(t)-\FDP_{A,1}(t)|$. With similar arguments as above, we can show that 
\begin{equation*}
|\FDP_1(t)-\FDP_{A,1}(t)|=O(|\bone^T\bF(\widetilde{\bW}_1-\bW)|/R(t)). 
\end{equation*}
The infeasible least squares estimator $\widetilde{\bW}_1$ is 
\begin{eqnarray*}
\widetilde{\bW}_1=(\bF^T\bF)^{-1}\bF^Tvec(\widetilde{\bX})&=&(\bF^T\bF)^{-1}\bF^T(vec(\bmu^{\star})+\bF\bW+\bepsilon)\\
&=&(\bF^T\bF)^{-1}\bF^Tvec(\bmu^{\star})+\bW+(\bF^T\bF)^{-1}\bF^T\bepsilon. 
\end{eqnarray*}
For the third term, note that $E(\bF^T\bF)^{-1}\bF^T\bepsilon=0$ because the mean of $\bepsilon$ is 0. More interestingly, 
\begin{equation*}
var((\bF^T\bF)^{-1}\bF^T\bepsilon)=(\bF^T\bF)^{-1}\bF^Tvar(\bepsilon)\bF(\bF^T\bF)^{-1}, 
\end{equation*}
but the columns of $\bF$ are orthogonal to $var(\bepsilon)$. Therefore, $var((\bF^T\bF)^{-1}\bF^T\bepsilon)=0$. Consequentially, we have shown that $(\bF^T\bF)^{-1}\bF^T\bepsilon=0$. 

Hence, 
\begin{equation*}
|\bone^T\bF(\widetilde{\bW}_1-\bW)|=|\bone^T(\sum_{k=1}^h\brho_k\brho_k^T)vec(\bmu^{\star})|\leq(pq)^{1/2}\|\sum_{k=1}^h\brho_k\brho_k^T\|\|vec(\bmu^{\star})\|=(pq)^{1/2}\|vec(\bmu^{\star})\|. 
\end{equation*}

Therefore, we have shown that 
\begin{equation*}
|\FDP_1(t)-\FDP_{A,1}(t)|=O_p((pq)^{\zeta}((pq)^{-1/2}\|vec(\bmu^{\star})\|)). 
\end{equation*}

Next we will show that 
\begin{equation*}
\|\widehat{\bSigma}_2\otimes\widehat{\bSigma}_1-\bSigma_2\otimes\bSigma_1\|=O_p(pq(n+m)^{-1/2}). 
\end{equation*}
By triangular inequality, 
\begin{eqnarray*}
&&\|\widehat{\bSigma}_2\otimes\widehat{\bSigma}_1-\bSigma_2\otimes\bSigma_1\|\\
&=&\|\widehat{\bSigma}_2\otimes\widehat{\bSigma}_1-\widehat{\bSigma}_2\otimes\bSigma_1+\widehat{\bSigma}_2\otimes\bSigma_1-\bSigma_2\otimes\bSigma_1\|\\
&\leq&\|\widehat{\bSigma}_2\otimes(\widehat{\bSigma}_1-\bSigma_1)\|+\|(\widehat{\bSigma}_2-\bSigma_2)\otimes\bSigma_1\|. 
\end{eqnarray*}
By the property of Kronecker product, the operator norm of the product of two matrices $\bA$ and $\bB$ will be the product of the largest eigenvalue of $\bA$ and the largest eigenvalue of $\bB$. Therefore, the last line is 
\begin{equation*}
\|\widehat{\bSigma}_2\|\times\|\widehat{\bSigma}_1-\bSigma_1\|+\|\widehat{\bSigma}_2-\bSigma_2\|\times\|\bSigma_1\|. 
\end{equation*}
We denote the $(i,j)$th element in $\widehat{\bSigma}_1$ as $\widehat{\bSigma}_{1,(i,j)}$, then 
\begin{eqnarray*}
\widehat{\bSigma}_{1,(i,j)}&=&\frac{1}{q}\sum_{s=1}^q\frac{1}{n+m-2}\Big[\sum_{l=1}^n(\bY_{l,(i,s)}-\frac{1}{n}\sum_{l=1}^n\bY_{l,(i,s)})(\bY_{l,(j,s)}-\frac{1}{n}\sum_{l=1}^n\bY_{l,(j,s)})/(\widehat{\sigma}_{is}\widehat{\sigma}_{js})\\
&&\quad\quad\quad\quad+\sum_{k=1}^m(\bZ_{k,(i,s)}-\frac{1}{m}\sum_{k=1}^m\bZ_{k,(i,s)})(\bZ_{k,(j,s)}-\frac{1}{m}\sum_{k=1}^m\bZ_{k,(j,s)})/(\widehat{\sigma}_{is}\widehat{\sigma}_{js})\Big]. 
\end{eqnarray*}
Note that 
\begin{eqnarray*}
&&\sum_{l=1}^n(\bY_{l,(i,s)}-\frac{1}{n}\sum_{l=1}^n\bY_{l,(i,s)})(\bY_{l,(j,s)}-\frac{1}{n}\sum_{l=1}^n\bY_{l,(j,s)})/(\widehat{\sigma}_{is}\widehat{\sigma}_{js})\\
&=&\sum_{l=1}^n\big[\bY_{l,(i,s)}-\bmu_{is}-(\frac{1}{n}\sum_{l=1}^n\bY_{l,(i,s)}-\bmu_{is})\big]\big[\bY_{l,(j,s)}-\bmu_{js}-(\frac{1}{n}\sum_{l=1}^n\bY_{l,(j,s)}-\bmu_{js})\big]/(\widehat{\sigma}_{is}\widehat{\sigma}_{js})\\
&=&\sum_{l=1}^n(\bY_{l,(i,s)}-\bmu_{is})(\bY_{l,(j,s)}-\bmu_{js})/(\widehat{\sigma}_{is}\widehat{\sigma}_{js})-\frac{1}{n}\sum_{l=1}^n(\bY_{l,(i,s)}-\bmu_{is})\sum_{l=1}^n(\bY_{l,(j,s)}-\bmu_{js})/(\widehat{\sigma}_{is}\widehat{\sigma}_{js}). 
\end{eqnarray*}
Note that 
\begin{equation*}
E[\sum_{l=1}^n(\bY_{l,(i,s)}-\bmu_{is})(\bY_{l,(j,s)}-\bmu_{js})/(\sigma_{is}\sigma_{js})]=n\bSigma_{1,(i,j)}, 
\end{equation*}
and 
\begin{equation*}
\frac{1}{n}E[\sum_{l=1}^n(\bY_{l,(i,s)}-\bmu_{is})/\sigma_{is}\sum_{l=1}^n(\bY_{l,(j,s)}-\bmu_{js})/\sigma_{js}]=\frac{1}{n}\sum_{l=1}^nE[(\bY_{l,(i,s)}-\bmu_{is})(\bY_{l,(j,s)}-\bmu_{js})/(\sigma_{is}\sigma_{js})]=\bSigma_{1,(i,j)}. 
\end{equation*}
Therefore, 
\begin{equation*}
\frac{1}{n-1}E(\sum_{l=1}^n(\bY_{l,(i,s)}-\frac{1}{n}\sum_{l=1}^n\bY_{l,(i,s)})(\bY_{l,(j,s)}-\frac{1}{n}\sum_{l=1}^n\bY_{l,(j,s)})/(\sigma_{is}\sigma_{js}))=\bSigma_{1,(i,j)}. 
\end{equation*}
On the other hand,
\begin{eqnarray*}
Var(\frac{1}{n-1}\sum_{l=1}^n(\bY_{l,(i,s)}-\frac{1}{n}\sum_{l=1}^n\bY_{l,(i,s)})(\bY_{l,(j,s)}-\frac{1}{n}\sum_{l=1}^n\bY_{l,(j,s)})/(\sigma_{is}\sigma_{js}))=O(n^{-1}). 
\end{eqnarray*}
Therefore, 
\begin{equation*}
\frac{1}{n-1}\sum_{l=1}^n(\bY_{l,(i,s)}-\frac{1}{n}\sum_{l=1}^n\bY_{l,(i,s)})(\bY_{l,(j,s)}-\frac{1}{n}\sum_{l=1}^n\bY_{l,(j,s)})/(\sigma_{is}\sigma_{js})=\bSigma_{1,(i,j)}+O_p(n^{-1/2}). 
\end{equation*}
Similarly, we can show that 
\begin{equation*}
\frac{1}{m-1}\sum_{k=1}^m(\bZ_{k,(i,s)}-\frac{1}{m}\sum_{k=1}^m\bZ_{k,(i,s)})(\bZ_{k,(j,s)}-\frac{1}{m}\sum_{k=1}^m\bZ_{k,(j,s)})/(\sigma_{is}\sigma_{js})=\bSigma_{1,(i,j)}+O_p(m^{-1/2})
\end{equation*}

Note that $\widehat{\sigma}_{ij}^2\sim(n+m-2)^{-1}\sigma_{ij}^2\chi_{n+m-2}^2$, Therefore, $\sigma_{ij}^2/\widehat{\sigma}_{ij}^2$ follows an inverse-chi-squared distribution with degrees of freedom as $n+m-2$ multiplied by a scalar $(n+m-2)$. Based on standard mean-variance analysis, we have
\begin{equation*}
\frac{\sigma_{ij}^2}{\widehat{\sigma}_{ij}^2}=\frac{n+m-2}{n+m-4}+O_p(\frac{n+m-2}{n+m-4}(n+m-6)^{-1/2}). 
\end{equation*}

Combining the two parts, for all $i$ and $j$, we have 
\begin{eqnarray*}
\widehat{\bSigma}_{1,(i,j)}&=&\frac{1}{q}\sum_{s=1}^q\frac{1}{n+m-2}\{(n-1)(\bSigma_{1,(i,j)}+O_p(n^{-1/2}))(1+O_p((n+m)^{-1/2})\\
&&\quad\quad+(m-1)(\bSigma_{1,(i,j)}+O_p(m^{-1/2}))(1+O_p((n+m)^{-1/2})\}\\
    &=&\bSigma_{1,(i,j)}+O_p((n+m)^{-1/2}). 
\end{eqnarray*}
By the property of matrix norms, 
\begin{equation*}
\|\widehat{\bSigma}_1-\bSigma_1\|\leq\|\widehat{\bSigma}_1-\bSigma_1\|_1=\max_{1\leq i\leq p}\sum_{j=1}^p|\widehat{\bSigma}_{1,(ij)}-\bSigma_{1,(ij)}|=O_p(p(n+m)^{-1/2}). 
\end{equation*}
Similarly, we have 
\begin{equation*}
\|\widehat{\bSigma}_2-\bSigma_2\|=O_p(q(n+m)^{-1/2}). 
\end{equation*}
By triangular inequality, we have 
\begin{equation*}
\|\widehat{\bSigma}_2\|\leq\|\bSigma_2\|+O_p(q(n+m)^{-1/2})\leq q+O_p(q(n+m)^{-1/2}). 
\end{equation*}
Therefore, 
\begin{equation*}
\|\widehat{\bSigma}_2\otimes\widehat{\bSigma}_1-\bSigma_2\otimes\bSigma_1\|=O_p(pq(n+m)^{-1/2}). 
\end{equation*}
By the triangular inequality, $|\theta_k-\widehat{\theta}|\geq||\theta_k-\theta_{k+1}|-|\theta_{k+1}-\widehat{\theta}_{k+1}||$. By Weyl's Theorem in Lemma 1, $|\theta_{k+1}-\widehat{\theta}_{k+1}|\leq\|\widehat{\bSigma}_2\otimes\widehat{\bSigma}_1-\bSigma_2\otimes\bSigma_1\|$. Therefore, on the event $\{\|\widehat{\bSigma}_2\otimes\widehat{\bSigma}_1-\bSigma_2\otimes\bSigma_1\|=O(pq(n+m)^{-1/2}\}$, $|\theta_k-\widehat{\theta}_{k+1}\|\geq d_{pq}-\|\widehat{\bSigma}_2\otimes\widehat{\bSigma}_1-\bSigma_2\otimes\bSigma_1\|\geq d_{pq}/2$ for sufficiently large $pq$. Similarly, we have $|\widehat{\theta}_{k-1}-\theta_k|\geq d_{pq}/2$. 

By $sin(\theta)$ Theorem in Lemma 1, we have
\begin{equation*}
\|\widehat{\brho}_k-\brho_k\|=O_p((n+m)^{-1/2}) 
\end{equation*}
for $k=1,\cdots,h$. 

By Weyl's Theorem in Lemma 1, 
\begin{equation*}
\sum_{k=1}^h|\widehat{\theta}_k-\theta_k|=O_p(hpq(n+m)^{-1/2}). 
\end{equation*}
Back to $\Delta_1$ in expression (\ref{delta1}), we have 
\begin{equation*}
|\Delta_1|=O_p(hpq(n+m)^{-1/2})+O_p(h(n+m)^{-1/2}(pq)^{1/2}\|vec(\bmu^{\star})\|). 
\end{equation*}
Under the assumption that $R(t)^{-1}=O_p((pq)^{-1+\zeta})$, we can show that 
\begin{equation*}
|\widehat{\FDP}_1(t)-\FDP_{A,1}(t)|=O_p((pq)^{\zeta}(h(n+m)^{-1/2}+(pq)^{-1/2}(\|\bmu^{\star}\|))). 
\end{equation*}
The proof is now complete. 

\indent{\bf Proof of Proposition 2:}

By the proof of Theorem 2 (i) in Fan and Han (2017), if we can show that 
\begin{equation*}
(pq)^{-1}\|cov(vec(\bepsilon))\|_F=O((pq)^{-\delta})
\end{equation*}
for some $\delta>0$, then the conclusion is correct. Note that 
\begin{equation*}
cov(vec(\bepsilon))=(\sum_{j=k_2+1}^q\xi_j\bgamma_j\bgamma_j^T)\otimes(\sum_{i=k_1+1}^p\lambda_i\bnu_i\bnu_i^T). 
\end{equation*}
Note that the eigenvalues of the Kronecker product are the product of the eigenvalues from the original two matrices, and the eigenvectors of the Kronecker product are the Kronecker product of the eigenvectors from the original two matrices. Therefore, we have 
\begin{equation*}
(\sum_{j=k_2+1}^q\xi_j\bgamma_j\bgamma_j^T)\otimes(\sum_{i=k_1+1}^p\lambda_i\bnu_i\bnu_i^T)=\sum_{j=k_2+1}^q\sum_{i=k_1+1}^p(\lambda_i\xi_j)(\bnu_i\otimes\bgamma_j)(\bnu_i\otimes\bgamma_j)^T. 
\end{equation*}
The Frobenius norm is invariant under orthonormal transformation. Therefore, 
\begin{equation*}
\|\sum_{j=k_2+1}^q\sum_{i=k_1+1}^p(\lambda_i\xi_j)(\bnu_i\otimes\bgamma_j)(\bnu_i\otimes\bgamma_j)^T\|_F^2=\sum_{j=k_2+1}^q\sum_{i=k_1+1}^p(\lambda_i\xi_j)^2=(\sum_{i=k_1+1}^p\lambda_i^2)(\sum_{j=k_2+1}^q\xi_j^2). 
\end{equation*}
By the conditions in Proposition 2, 
\begin{equation*}
(pq)^{-1}\|cov(vec(\bepsilon)\|_F=(pq)^{-1}(\sum_{i=k_1+1}^p\lambda_i^2)^{1/2}(\sum_{j=k_2+1}^q\xi_j^2)^{1/2}=O(p^{-\delta_1})O(q^{-\delta_2}). 
\end{equation*}
The conclusion is correct. Following the proof of Proposition 1, we can show the conclusion.

\indent{\bf Proof of Theorem 2: }

Based on model (\ref{vecx}), we consider a least squares estimator for $vec(\bW)$:
\begin{eqnarray*}
\widehat{vec(\bW)}&=&[(\bD^T\otimes\bC)^T(\bD^T\otimes\bC)]^{-1}(\bD^T\otimes\bC)^Tvec(\bX)\\
    &=&[(\bD\bD^T)\otimes(\bC^T\bC)]^{-1}(\bD^T\otimes\bC)^Tvec(\bX)\\
    &=&[diag(\xi_1^{-1},\cdots,\xi_{k_2}^{-1})\otimes diag(\lambda_1^{-1},\cdots,\lambda_{k_1}^{-1})](\bD^T\otimes\bC)^Tvec(\bX). 
\end{eqnarray*}
Correspondingly, 
\begin{eqnarray*}
(\bD^T\otimes\bC)\widehat{vec(\bW)}&=&[(\xi_1^{-1/2}\bgamma_1,\cdots,\xi_{k_2}^{-1/2}\bgamma_{k_2})\otimes(\lambda_1^{-1/2}\bnu_1,\cdots,\lambda_{k_1}^{-1/2}\bnu_{k_1})]\\
&&\times[(\xi_1^{1/2}\bgamma_1,\cdots,\xi_{k_2}^{1/2}\bgamma_{k_2})^T\otimes(\lambda_1^{1/2}\bnu_1,\cdots,\lambda_{k_1}^{1/2}\bnu_{k_1})^T]vec(\bX)\\
&=&[(\sum_{i=1}^{k_2}\bgamma_i\bgamma_i^T)\otimes(\sum_{j=1}^{k_1}\bnu_j\bnu_j^T)]vec(\bX). 
\end{eqnarray*}

In $\bD^T\otimes\bC$, each column is $\sqrt{\xi_l\lambda_k}\bgamma_l\otimes\bnu_k$, for $l=1,\cdots,k_2$ and $k=1,\cdots, k_1$. Similar to the proof of Theorem 1, we can show that there exist positive constants $C_1, C_2, C_3, C_4$ such that 
\begin{eqnarray}\label{fourterm}
|\widehat{\FDP}_2(t)-\FDP_{A,2}(t)|&\leq&\frac{1}{R(t)}\Big[C_1\sum_{l=1}^{pq}|\widehat{\bb}_l^T\widehat{\bW}-\bb_l^T\widetilde{\bW}_2|+C_2\sum_{l=1}^{pq}(\|\widehat{\bb}_l\|^2-\|\bb_l\|^2)\nonumber\\
&&+C_3\sum_{l=1}^{pq}|\bb_l^T(\widetilde{\bW}_2-\widetilde{\bW}_1)|+C_4|\bone^T(\bD^T\otimes\bC)(\widetilde{\bW}_1-\bW)|\Big]. 
\end{eqnarray}

We will analyze each term in the last line. For the second term in (\ref{fourterm}), we have
\begin{eqnarray}\label{triplestar}
&&\sum_{l=1}^{pq}[\|\widehat{\bb}_l\|^2-\|\bb_l\|^2]\nonumber\\
&=&\sum_{j=1}^q\sum_{i=1}^p|\sum_{k=1}^{k_1}\sum_{l=1}^{k_2}(\sqrt{\widehat{\xi}_l\widehat{\lambda}_k}\widehat{\bgamma}_{li}\widehat{\bnu}_{kj})^2-\sum_{k=1}^{k_1}\sum_{l=1}^{k_2}(\sqrt{\xi_l\lambda_k}\bgamma_{li}\bnu_{kj})^2|\nonumber\\
&=&\sum_{j=1}^q\sum_{i=1}^p|\sum_{k=1}^{k_1}\sum_{l=1}^{k_2}(\widehat{\xi}_l\widehat{\lambda}_k-\xi_l\lambda_k)\widehat{\bgamma}_{li}^2\widehat{\bnu}_{kj}^2+\sum_{k=1}^{k_1}\sum_{l=1}^{k_2}\xi_l\lambda_k(\widehat{\bgamma}_{li}^2\widehat{\bnu}_{kj}^2-\bgamma_{li}^2\bnu_{kj}^2)|\nonumber\\
&\leq&\sum_{k=1}^{k_1}\sum_{l=1}^{k_2}|\widehat{\xi}_l\widehat{\lambda}_k-\xi_l\lambda_k|(\sum_{j=1}^q\sum_{i=1}^p\widehat{\bgamma}_{li}^2\widehat{\bnu}_{kj}^2)+\sum_{k=1}^{k_1}\sum_{l=1}^{k_2}\xi_l\lambda_k\sum_{j=1}^q\sum_{i=1}^p|\widehat{\bgamma}_{li}^2\widehat{\bnu}_{kj}^2-\bgamma_{li}^2\bnu_{kj}^2|. 
\end{eqnarray}
For the second term in (\ref{triplestar}),
\begin{eqnarray*}
&&\sum_{j=1}^q\sum_{i=1}^p|\widehat{\bgamma}_{li}^2\widehat{\bnu}_{kj}^2-\bgamma_{li}^2\bnu_{kj}^2|\\
&=&\sum_{j=1}^q\sum_{i=1}^p|\widehat{\bgamma}_{li}\widehat{\bnu}_{kj}+\bgamma_{li}\bnu_{kj}|\times |\widehat{\bgamma}_{li}\widehat{\bnu}_{kj}-\bgamma_{li}\bnu_{kj}|\\
&\leq&(\sum_{j=1}^q\sum_{i=1}^p|\widehat{\bgamma}_{li}\widehat{\bnu}_{kj}+\bgamma_{li}\bnu_{kj}|^2)^{1/2}(\sum_{j=1}^q\sum_{i=1}^p(\widehat{\bgamma}_{li}\widehat{\bnu}_{kj}-\bgamma_{li}\bnu_{kj})^2)^{1/2}\\
&\leq&[2\sum_{j=1}^q\sum_{i=1}^p(\widehat{\bgamma}_{li}^2\widehat{\bnu}_{kj}^2+\bgamma_{li}^2\bnu_{kj}^2)]^{1/2}\times \|\widehat{\bgamma}_l\otimes\widehat{\bnu}_k-\bgamma_l\otimes\bnu_k\|\\
&=&2 \|\widehat{\bgamma}_l\otimes\widehat{\bnu}_k-\bgamma_l\otimes\bnu_k\|. 
\end{eqnarray*}
Therefore, for some positive constant, 
\begin{equation*}
\sum_{l=1}^{pq}\big[\|\widehat{\bb}_l^2-\|\bb_l\|^2\big]\leq C(\sum_{k=1}^{k_1}\sum_{l=1}^{k_2}|\widehat{\xi}_l\widehat{\lambda}_k-\xi_l\lambda_k|+\xi_l\lambda_k\|\widehat{\bgamma}_l\otimes\widehat{\bnu}_k-\bgamma_l\otimes\bnu_k\|). 
\end{equation*}
For the first term in (\ref{triplestar}), 
\begin{eqnarray*}
&&\sum_{k=1}^{k_1}\sum_{l=1}^{k_2}|\widehat{\xi}_l\widehat{\lambda}_k-\xi_l\lambda_k|\\
&=&\sum_{k=1}^{k_1}\sum_{l=1}^{k_2}|\widehat{\xi}_l\widehat{\lambda}_k-\xi_l\widehat{\lambda}_k+\xi_l\widehat{\lambda}_k-\xi_l\lambda_k|\\
&\leq&\sum_{k=1}^{k_1}\sum_{l=1}^{k_2}(|\widehat{\xi}_l-\xi_l|\widehat{\lambda}_k+\xi_l|\widehat{\lambda}_k-\lambda_k|)\\
&=&\sum_{l=1}^{k_2}|\widehat{\xi}_l-\xi_l|\sum_{k=1}^{k_1}\widehat{\lambda}_k+\sum_{l=1}^{k_2}\xi_l\sum_{k=1}^{k_1}|\widehat{\lambda}_k-\lambda_k|. 
\end{eqnarray*}
In the proof of Theorem 1, we have shown $\|\widehat{\bSigma}_1-\bSigma_1\|=O_p(p(n+m)^{-1/2})$ and $\|\widehat{\bSigma}_2-\bSigma_2\|=O_p(q(n+m)^{-1/2})$. 
Note that in the last line, $\sum_{l=1}^{k_2}|\widehat{\xi}_l-\xi_l|=O_p(k_2q(n+m)^{-1/2})$, and $\sum_{k=1}^{k_1}|\widehat{\lambda}_k-\lambda_k|=O_p(k_1p(n+m)^{-1/2})$. Furthermore, we have $\sum_{l=1}^{k_2}\xi_l\leq q$ and 
\begin{equation*}
\sum_{k=1}^{k_1}\widehat{\lambda}_k=\sum_{k=1}^{k_1}(\widehat{\lambda}_k-\lambda_k+\lambda_k)=O_p(k_1p(n+m)^{-1/2})+p. 
\end{equation*}

Next, for the first term in (\ref{triplestar}), we can show that 
\begin{equation*}
\sum_{l=1}^{pq}|\widehat{\bb}_l^T\widehat{\bW}-\bb_l^T\widetilde{\bW}_2|\\
\leq\sqrt{pq}\|(\sum_{i=1}^{k_2}\widehat{\bgamma}_i\widehat{\bgamma}_i^T)\otimes(\sum_{j=1}^{k_1}\widehat{\bnu}_j\widehat{\bnu}_j^T)-(\sum_{i=1}^{k_1}\bgamma_i\bgamma_i^T)\otimes(\sum_{j=1}^{k_1}\bnu_j\bnu_j^T)\|\|vec(\bX)\|. 
\end{equation*}
In the last line, by the triangular inequality and the property of Kronecker product on the eigenvalues, we have
\begin{eqnarray*}
&&\|(\sum_{i=1}^{k_2}\widehat{\bgamma}_i\widehat{\bgamma}_i^T)\otimes(\sum_{j=1}^{k_1}\widehat{\bnu}_j\widehat{\bnu}_j^T)-(\sum_{i=1}^{k_1}\bgamma_i\bgamma_i^T)\otimes(\sum_{j=1}^{k_1}\bnu_j\bnu_j^T)\|\\
&\leq&\|(\sum_{i=1}^{k_2}\widehat{\bgamma}_i\widehat{\bgamma}_i^T)\otimes(\sum_{j=1}^{k_1}\widehat{\bnu}_j\widehat{\bnu}_j^T-\sum_{j=1}^{k_1}\bnu_j\bnu_j^T)\|+\|(\sum_{i=1}^{k_2}\widehat{\bgamma}_i\widehat{\bgamma}_i^T-\sum_{i=1}^{k_2}\bgamma_i\bgamma_i^T)\otimes(\sum_{j=1}^{k_1}\bnu_j\bnu_j^T\|\\
&=&\|(\sum_{i=1}^{k_2}\widehat{\bgamma}_i\widehat{\bgamma}_i^T)\|\times\|\sum_{j=1}^{k_1}\widehat{\bnu}_j\widehat{\bnu}_j^T-\sum_{j=1}^{k_1}\bnu_j\bnu_j^T\|+\|\sum_{i=1}^{k_2}\widehat{\bgamma}_i\widehat{\bgamma}_i^T-\sum_{i=1}^{k_2}\bgamma_i\bgamma_i^T\|\times\|\sum_{j=1}^{k_1}\bnu_j\bnu_j^T\|\\
&=&\|\sum_{j=1}^{k_1}\widehat{\bnu}_j\widehat{\bnu}_j^T-\sum_{j=1}^{k_1}\bnu_j\bnu_j^T\|+\|\sum_{i=1}^{k_2}\widehat{\bgamma}_i\widehat{\bgamma}_i^T-\sum_{i=1}^{k_2}\bgamma_i\bgamma_i^T\|\\
&\leq&2\sum_{j=1}^{k_1}\|\widehat{\bnu}_j-\bnu_j\|+2\sum_{i=1}^{k_2}\|\widehat{\bgamma}_i-\bgamma_i\|. 
\end{eqnarray*}
By $\sin(\theta)$ Theorem, we can show that the last line is $O_p(k_1(n+m)^{-1/2})+O_p(k_2(n+m)^{-1/2})$. 

For the third term in (\ref{fourterm}), we have 
\begin{eqnarray*}
\sum_{l=1}^{pq}|\bb_l^T(\widetilde{\bW}_2-\widetilde{\bW}_1)|
   &\leq&\bone^T|(\sum_{i=1}^{k_2}\bgamma_i\bgamma_i^T)\otimes(\sum_{j=1}^{k_1}\bnu_j\bnu_j^T)(vec(\bX)-vec(\widetilde{\bX}))|\\
   &\leq&\sqrt{pq}\|(\sum_{i=1}^{k_2}\bgamma_i\bgamma_i^T)\otimes(\sum_{j=1}^{k_1}\bnu_j\bnu_j^T)\|\times\|vec(\bX)-vec(\widetilde{\bX})\|\\
   &=&\sqrt{pq}\|vec(\bX)-vec(\widetilde{\bX})\|. 
\end{eqnarray*}
In the proof of Theorem 1, we have shown that $\|vec(\bX)-vec(\widetilde{\bX})\|=O_p((n+m-2)^{-1/2}(\|vec(\bmu^{\star})\|+(pq)^{1/2}))$. 

For the last term in (\ref{fourterm}), similar to the proof of Theorem 1, we can show that $|\bone^T(\bD^T\otimes\bC)(\widetilde{\bW}_1-\bW))|=(pq)^{1/2}\|vec(\bmu^{\star})\|$. Combining the four terms in (\ref{fourterm}), we can show that 
\begin{equation*}
|\widehat{\FDP}_2(t)-\FDP_{A,2}(t)|=O_p((pq)^{\zeta}(k_1k_2(n+m)^{-1}+(k_1+k_2)(n+m)^{-1/2}+(pq)^{-1/2}\|vec(\bmu^{\star})\|). 
\end{equation*}
The proof is now complete. 

\vspace{0.3in}

\noindent{\bf Acknowledgement}

We would like to thank Dr. Yanhui Xu for her early assistant on the numerical studies.

\end{document}